\documentclass[twocolumn]{aastex631}

\received{December 1, 2022}
\accepted{\today}

\shorttitle{Calibrating UV-Based Estimates of Supermassive Black Hole Masses}
\shortauthors{Dix et al.}
\usepackage{amsmath}

\usepackage{bold-extra}
\usepackage{hyperref}
\usepackage{xcolor}
\newcommand{\angstrom}{\textup{\AA}}
\newcommand{\mbh}{$M_{\rm BH}$~}
\newcommand{\hb}{H$\beta$~}
\newcommand{\kms}{km s$^{-1}$}

\begin{document}


\title{Gemini Near Infrared Spectrograph - Distant Quasar Survey: Prescriptions for Calibrating UV-Based Estimates of Supermassive Black Hole Masses in High-Redshift Quasars}

\email{cooperdix@my.unt.edu}

\author[0000-0003-0192-1840]{Cooper Dix}
\affil{Department of Physics, University of North Texas, Denton, TX, 76203, USA \\}

\author[0000-0001-8406-4084]{Brandon Matthews}
\affil{Department of Physics, University of North Texas, Denton, TX, 76203, USA \\}

\author[0000-0003-4327-1460]{Ohad Shemmer}
\affil{Department of Physics, University of North Texas, Denton, TX, 76203, USA \\}

\author[0000-0002-1207-0909]{Michael S. Brotherton}
\affil{Department of Physics and Astronomy, University of Wyoming, Laramie, WY, 82071, USA \\}

\author{Adam D. Myers}
\affil{Department of Physics and Astronomy, University of Wyoming, Laramie, WY, 82071, USA \\}

\author[0000-0003-1562-5188]{I. Andruchow}
\affil{Facultad de Ciencias Astron\'omicas y Geosicas, Universidad Nacional de La Plata, Paseo del Bosque, B1900FWA, La Plata, Argentina}
\affil{Instituto Argentino de Radioastronom\'{\i}a, CONICET-CICPBA-UNLP, CC5 (1897) Villa Elisa, Prov. de Buenos Aires, Argentina}

\author[0000-0002-0167-2453]{W. N. Brandt}
\affil{Department of Astronomy and Astrophysics, The Pennsylvania State University, University Park, PA 16802, USA}
\affiliation{Institute for Gravitation and the Cosmos, The Pennsylvania State University, University Park, PA 16802, USA}
\affiliation{Department of Physics, 104 Davey Lab, The Pennsylvania State University, University Park, PA 16802, USA}

\author{Gabriel A. Ferrero}
\affil{Facultad de Ciencias Astron\'omicas y Geofísicas, Universidad Nacional de La Plata, Paseo del Bosque, B1900FWA La Plata, Argentina}
\affil{Instituto de Astrof\'isica de La Plata, CONICET–UNLP, CCT La Plata, Paseo del Bosque, B1900FWA La Plata, Argentina}

\author[0000-0003-1245-5232]{Richard Green}
\affil{Steward Observatory, University of Arizona, 933 N Cherry Ave, Tucson, AZ 85721, USA}

\author{Paulina Lira}
\affil{Departmento de Astronom\'{\i}a, Universidad de Chile, Casilla 36D, Santiago, Chile}


\author[0000-0002-7092-0326]{Richard M. Plotkin}
\affil{Department of Physics, University of Nevada, Reno, NV 89557, USA}
\affil{Nevada Center for Astrophysics, University of Nevada, Las Vegas, NV 89154, USA}

\author[0000-0002-1061-1804]{Gordon T. Richards}
\affil{Department of Physics, Drexel University, 32 S. 32nd Street, Philadelphia, PA 19104, USA}

\author[0000-0001-7240-7449]{Donald P. Schneider}
\affil{Department of Astronomy and Astrophysics, The Pennsylvania State University, University Park, PA 16802, USA}
\affiliation{Institute for Gravitation and the Cosmos, The Pennsylvania State University, University Park, PA 16802, USA}

\begin{abstract}
The most reliable single-epoch supermassive black hole mass ($M_{\rm BH}$) estimates in quasars are obtained by using the velocity widths of low-ionization emission lines, typically the H$\beta$~$\lambda4861$ line.
Unfortunately, this line is redshifted out of the optical band at $z\approx1$, leaving \mbh estimates to rely on proxy rest-frame ultraviolet (UV) emission lines, such as C~{\sc iv}~$\lambda1549$ or Mg~{\sc ii}~$\lambda2800$, which contain intrinsic challenges when measuring resulting in uncertain \mbh estimates.
In this work, we aim at correcting $M_{\rm BH}$ estimates derived from the C~{\sc iv} and Mg~{\sc ii} emission lines based on estimates derived from the H$\beta$ emission line.
We find that employing the equivalent width of C~{\sc iv} in deriving \mbh estimates based on Mg~{\sc ii} and C~{\sc iv} provides values that are closest to those obtained from H$\beta$.
We also provide prescriptions to estimate \mbh values when only C~{\sc iv}, only Mg~{\sc ii}, and both C~{\sc iv} and Mg~{\sc ii} are measurable.
We find that utilizing both emission lines, where available, reduces the scatter of UV-based \mbh estimates by $\sim15\%$ when compared to previous studies.
Lastly, we discuss the potential of our prescriptions to provide more accurate and precise estimates of \mbh given a much larger sample of quasars at \hbox{$3.20 \lesssim z \lesssim 3.50$}, where both Mg~{\sc ii} and H$\beta$ can be measured in the same near-infrared spectrum.
\end{abstract}

\section{Introduction} \label{sec:intro}

A persisting point of interest in astrophysics today is understanding the co-evolution of supermassive black holes (SMBHs) and their host galaxies through cosmic time \citep[e.g.,][]{Hop06,DiM08,BaY11,Car18,Chen20,Suh20}.
A fundamental ingredient in this research area is the SMBH mass ($M_{\rm BH}$).
Over the past four decades, several methods have been employed for obtaining \mbh values in galaxies \citep[such as stellar kinematics, masers, interferometry and spectrophotometric monitoring campaigns of active galaxies, e.g.,][] {Fer00,Geb00,Gre05,Gul09,Gre10,Shen15,Gri19,Gra22}.
Overall, the masses obtained from these methods are consistent with each other but deriving \mbh values in active galactic nuclei (AGN) have the best prospects of obtaining the SMBH mass function through cosmic time given the large luminosities of such sources and their observable mass indicators at all accessible redshifts \citep[e.g.,][]{Kel10,KaM12,SaK12,TaN12}.

The \mbh values for AGN, or quasars, are usually determined through measurements of broad emission lines in the optical band.
Specifically, following the virial assumption \citep[see,][]{PaW99}, we use measurements of the size of the broad emission line region (BELR), $R_{\rm BELR}$, and the velocity width of an emission line stemming from the BELR, $\Delta V$, in order to estimate $M_{\rm BH}$ for AGN.
Of these terms, estimating the value of $R_{\rm BELR}$ becomes the most pertinent for reliable estimates of $M_{\rm BH}$.

Ideally, measurements of $R_{\rm BELR}$ are derived from reverberation mapping (RM) of AGN or quasars, which uses time lags between continuum fluctuations and photoionized BELR emission line fluctuations to determine the size of the BELR \citep[e.g.,][]{BaM82,Pet93,Pan14}.
To date, \mbh has been measured successfully using RM campaigns for $\approx150$ quasars primarily with the \hb $\lambda4861$ emission line \citep[e.g.,][]{Bar15,BaK15,Gri17,Du18a,Hu21,Bao22,U22}.
One of the most important findings from these RM campaigns is the BELR size-luminosity ($R - L$) relation, where $R_{\rm BELR} \propto L^{\alpha}$ with $\alpha \sim 0.5$, in agreement with expectations from photoionization theory \citep[e.g.,][]{Laor98,Kas00,Kas05,Ben09,Ben13}.

Since RM campaigns are currently impractical for \mbh measurements in $\approx10^{6}$ of known quasars \citep[e.g.,][]{Shen15}, \citet[][]{VaP06} have proposed that the $R - L$ relation, in conjuction with the virial assumption, allows one to \textit{estimate} single epoch (SE) \mbh values by substituting the continuum luminosity for $R_{\rm BELR}$.
Estimates of \mbh values for $\approx10^{5}$ quasars have been obtained in this fashion during the past two decades \citep[e.g.,][]{Shen11,Rak20,Wu&Shen22}.

Nevertheless, estimating \mbh values using the SE method faces additional challenges, particularly at high redshift.
First, the most reliable SE indicator for \mbh is obtained from spectroscopic measurements of low-ionization emission lines such as the \hb line, and at $z\gtrsim1$, this line is shifted into the less accessible near-infrared (NIR) band.
Second, recent Super-Eddington Accreting Massive Black Hole (SEAMBH) and Sloan Digital Sky Survey-RM campaigns discovered many highly accreting objects that lie below the $R - L$ relation \citep[e.g.,][]{Du18a,Fon20}, suggesting that an additional correction to account for accretion rate is warranted for SE \mbh estimates.

To overcome the first of these, SE \mbh estimates using other prominent emission lines have been calibrated against H$\beta$-based \mbh estimates in the nearby universe. 
The two most common emission lines that are used for such calibrations are Mg~{\sc ii} $\lambda\lambda2798,2803$ \citep[e.g.,][]{MaD04,VaO09,Zuo15,Woo18,Le20} and C~{\sc iv} $\lambda1549$ \citep[e.g.,][]{VaP06,Ass11,Run13,Bro15,Park17,Coa17,Sun18,Dal20}. 
However, these emission lines have yielded relatively fewer successful \mbh measurements through RM campaigns \citep[e.g.,][]{Cac15,Shen16b,Lir18,Gri19,Hoo19,Hom20,Kas21}, and each of these line profiles contains its own intrinsic measurement challenges \citep[e.g.][]{VaW01,BaL05}. 
To address the second challenge, \citet{Du19} have proposed to include a correction to the $R - L$ relationship based on the Fe~{\sc ii} emission blend flanking the H$\beta$ emission line, which is known to be an accretion-rate indicator.
Recently, \citet{Mai22} implemented such a correction and found that \mbh estimates in highly accreting sources are overestimated.

In this work, we utilize a large spectroscopic inventory of high-redshift quasars that allows us to obtain the most reliable \mbh estimates using rest-frame ultraviolet (UV) emission lines.
Our inventory includes high quality measurements of the H$\beta$, Fe~{\sc ii}, Mg~{\sc ii}, and C~{\sc iv} emission lines, which allows us to implement two separate accretion-rate based corrections to the estimated \mbh value while investigating the effects of using different BELR velocity width measurements. 

This paper is organized as follows. In Section \ref{sec:sample}, we describe our sample and data analysis. In Section \ref{sec:analysis}, we present the results of multiple regression analyses used for obtaining prescriptions for reliable \mbh estimates at high redshift. In Section \ref{sec:discussion} we discuss our results and in Section \ref{sec:conclusions} we present our conclusions. Throughout this paper, we compute luminosity distances using $H_{0} = 70$ km s$^{-1}$ Mpc$^{-1}$, $\Omega_{\rm M} = 0.3$, and $\Omega _{\Lambda} = 0.7$ \citep[e.g.,][]{Spe07}.

\section{Sample Selection and Measurements} \label{sec:sample}

Our sample is drawn from the Gemini Near Infrared Spectrograph - Distant Quasar Survey \citep[GNIRS-DQS;][hereafter Paper I]{Mat23}.
%
Details of this survey, the data quality, and all spectral fits performed for each source are described in \citet[][hereafter M21]{Mat21} and Paper I.
Briefly, GNIRS-DQS utilizes spectroscopy from the GNIRS instrument \citep{2006SPIE.6269E..4CE} in the $\sim 0.8-2.5 \mu$m wavelength band at a spectral resolution of $R \sim 1100$ to construct the largest, uniform rest-frame optical spectral inventory for high-redshift quasars (see, M21).
The GNIRS-DQS sources were selected from all the Sloan Digital Sky Survey \citep[SDSS;][]{Yor00} quasars \citep{Lyke20} having $m_{i}$ values up to $ \sim 19.0$ that lie in the redshift intervals \hbox{$1.55 \lesssim z \lesssim 1.65$}, \hbox{$2.10 \lesssim z \lesssim 2.40$}, and \hbox{$3.20 \lesssim z \lesssim 3.50$}; these redshift intervals assure that the H$\beta$ spectral region is covered in either the $J$, $H$, or $K$ bands. 

From all 260 GNIRS-DQS sources, we were able to practically measure C~{\sc iv} emission-line properties for 177 sources from their respective SDSS spectra. 
Typically, this emission line cannot be measured reliably in both broad absorption line (BAL) quasars and radio-loud quasars (RLQs).\footnote{We define radio loud quasars as sources having radio-loudness values of $R > 100$ \citep[where $R$ is the ratio of the flux densities at 5 GHz and $4400$ \AA ;][Paper I]{Kel89}.}
Specifically, the C~{\sc iv} emission line is difficult to measure in BAL quasars due to BAL troughs often impacting the emission-line profile.
Therefore, all 65 BAL quasars from the GNIRS-DQS sample were removed during our C~{\sc iv}-based \mbh estimate analysis.
Additionally, since our analysis involves measurements of the rest-frame equivalent width (EW) of the C~{\sc iv} emission line, we further removed 16 RLQs from the sample. 
This was done in order to avoid potential dilution of the C~{\sc iv} emission line by continuum emission originating in the radio jets. 
We note that one of the BAL quasars we removed, \hbox{SDSS J114705.24$+$083900.6}, is also radio loud. 
Finally, we removed two sources, SDSS \hbox{J073132.18+461347.0} and SDSS \hbox{J141617.38+264906.1}, for which we were unable to measure the C~{\sc iv} emission line reliably from their SDSS spectra.
Specifically, the SDSS spectrum of \hbox{J073132.18+461347.0} contains pixels with highly uncertain flux densities over a large portion of the C~{\sc iv} profile, while the spectrum of SDSS \hbox{J141617.38+264906.1} suffers from significant narrow line absorption, directly affecting the C~{\sc iv} profile, preventing us from obtaining a reliable line profile for both of these sources.
The remaining sample of 177 non-BAL, non-RL sources with reliable C~{\sc iv} measurements was used in the C~{\sc iv}-based \mbh estimate analysis below.

The GNIRS spectra provide Mg~{\sc ii} measurements for 99 of the GNIRS-DQS sources (see, Paper I): only 70 of these sources also have corresponding C~{\sc iv} measurements following the removal of 22 BAL quasars and seven RLQs.
From these 99 quasars, 65 (47 with reliable C~{\sc iv} measurements) lie in the redshift range of \hbox{$2.10 \lesssim z \lesssim 2.40$}, and 34 (23 with reliable C~{\sc iv} measurements) lie at \hbox{$3.20 \lesssim z \lesssim 3.50$}.
In both of these redshift ranges Mg~{\sc ii} and H$\beta$ are covered in the same spectrum, however, in the latter range Mg~{\sc ii} has the highest signal-to-noise (S/N) ratio \citep[see below, and cf.][]{Zuo15}.

Furthermore, we were able to measure the Mg~{\sc ii} profile in the SDSS spectra that adequately covered that emission line in 179 of the GNIRS-DQS sources: 34 and 13 of these sources do not have reliable C~{\sc iv} measurements given that these are BAL quasars and RLQs, respectively.
From this sample of 179 quasars, 53 sources had a measurable Mg~{\sc ii} profile in both the SDSS and the GNIRS-DQS spectra.
When combining all available Mg~{\sc ii} measurements, either from SDSS or GNIRS-DQS or both, we compiled a total sample of 225 sources: 47, 16, and 2 of these sources do not have reliable C~{\sc iv} measurements given that these are BAL quasars, RLQs, or sources without adequate C~{\sc iv} measurements, respectively. 

\subsection{Fitting the SDSS Spectra}

The fitting procedure performed for the SDSS spectra in this work follows the methodology described in  \citet{2020ApJ...893...14D}.
In short, this was done utilizing a local linear continuum and two Gaussians for each broad emission line.
We find that fitting two Gaussians to the entire profile of the C~{\sc iv} and Mg~{\sc ii} emission lines is sufficient given the S/N of $\sim40$ per pixel across both the SDSS and GNIRS spectra.
The Fe~{\sc ii} and Fe~{\sc iii} emission complex that blends with the Mg~{\sc ii} emission line was modeled with the empirical template of \citet{VaW01}.
This template was chosen for consistency between the Mg~{\sc ii} fits presented in this work and those from Paper I.
While this template does not account for Fe emission underlying the Mg~{\sc ii} emission line, previous studies conclude this template overestimates the Mg~{\sc ii} full width at half maximum (FWHM) intensity by up to $\sim20\%$ \citep[e.g.,][]{2011ApJ...739...56D,2020ApJ...898..105O,2020ApJ...905...51S,2021ApJ...923..262Y}.
Overall, given the uncertainties of the Mg~{\sc ii} emission line measurements in the GNIRS spectra of our sources (see, Paper I) and the intrinsic uncertainty of SE \mbh estimates (see, Section \ref{sec:discussion}), we expect any uncertainties associated with adopting this template to be modest for this analysis.
This template was broadened with a Gaussian kernel having a FWHM intensity that was free to vary up to \hbox{10000 \kms} and was determined based on a least squares analysis of each fitted region.

\indent The Gaussians were constrained such that the flux density would lie between 0 and twice the value of the peak of the respective emission line and the FWHM was restricted to lie within 0 and \hbox{15000 \kms}. 
The peaks of these Gaussians were also constrained to lie within $\pm1500$ \kms ~of the rest-frame wavelength of the peak of the emission line based on the systemic redshift from Paper I.
After the initial fitting was performed for each region, we visually inspected the fit to see if more lenient constraints with interactive fitting were warranted.

We excluded BAL and RLQs throughout this work in order to avoid potentially large uncertainties in the properties of the C~{\sc iv} emission line, as described above.
However, our derived prescriptions should be applicable to any quasar, given that a C~{\sc iv} emission line can be measured reliably in its spectrum.

Spectral properties stemming from these fits are reported in Table \ref{tab:c4} for C~{\sc iv} and Mg~{\sc ii}. 
In this Table, Column (1) reports the source's SDSS designation.
Columns (2), (3), (4), (5), and (6) list the FWHM, mean absolute deviation (MAD; described below), line dispersion ($\sigma_{\rm line}$), rest-frame EW, and the observed-frame wavelength of the emission-line peak, $\lambda_{\rm peak}$, respectively, for C~{\sc iv}. Columns (7), (8), (9), (10), and (11) list the same spectral properties for the Mg~{\sc ii} emission line.

\subsection{Measurements and Error}\label{sec:meas}

For each emission-line profile in either the GNIRS or SDSS spectra, we measured the values of the $\sigma_{\rm line}$ and MAD. The line dispersion is defined by\\
\begin{equation}\label{eq:3}
\sigma_{\rm line} = \bigg[ \frac{\int (\lambda - \lambda_{0})^{2}P(\lambda)d\lambda}{\int P(\lambda)d\lambda}\bigg]^{1/2}
\end{equation}\\
where $\lambda_{0}$ is the line centroid and $P(\lambda)$ is the emission-line profile. The MAD is defined as\\
\begin{equation}\label{eq:4}
\left. {\rm MAD} = \int |\lambda - \lambda_{\rm med}|P(\lambda)d\lambda \middle/ \int P(\lambda)d\lambda \right. ,
\end{equation}\\
where $\lambda_{\rm med}$ is the median wavelength of the emission-line profile, first suggested in \citet{Den16} as an appropriate representation for the emission-line width.
For each emission-line profile in the GNIRS spectra, we obtained the FWHM, EW, and observed-frame wavelength of the peak emission from Paper I.

We present three different values for the velocity widths (FWHM, MAD, $\sigma_{\rm line}$) due to the uncertainties inherent in using FWHM, the most popular of these parameters \citep[see,][]{Park17,Dal20,Le20}.
While $\sigma_{\rm line}$ is a dependable measurement to describe the emission-line velocity width, \citet{Den16} suggest that MAD provides a more accurate estimate of this quantity for low-quality data.
Overall, we recognize that the best virial velocity width indicator is debatable, therefore, we provide calibrations for the \mbh estimates utilizing all of these parameters.

We have also derived the monochromatic luminosities, $L_{1350}$ and $L_{3000}$, by measuring the continuum flux densities, at rest-frame \hbox{$\lambda1350$ \AA} and \hbox{$\lambda3000$ \AA }, respectively, and employing our chosen cosmology.
All the flux densities and monochromatic luminosities at rest-frame \hbox{$\lambda5100$ \AA\ }($L_{5100}$) used in this work were obtained from Paper I.
The flux calibration for the GNIRS-DQS spectra is extensively discussed in M21.
In our $z < 1.65$ sources, the flux density at rest-frame wavelength $3000$ \AA\ was not measurable in the GNIRS-DQS spectrum due to this wavelength range falling blueward of the $J$ band.
In these cases, the flux density was determined by extrapolating from the flux density at rest-frame wavelength \hbox{$5100$~\AA\ } ~using the canonical quasar optical-UV continuum of the form $f_{\nu} \propto \nu^{-0.5}$ \citep[e.g.,][]{RaS80,Van01}.
Similarly, there are SDSS spectra that do not have a reliable flux density value for the rest-frame wavelength \hbox{$1350$~\AA} due to low S/N at the blue end of the SDSS spectrum.
In these cases, we employed the same model as described above extrapolating from the flux density at rest-frame \hbox{$1450$~\AA.}

The uncertainties for all emission line measurements reported in Table 1 were determined by following the methods described in M21 and Paper I. Briefly, we created mock spectra that introduced random Gaussian noise to the original spectra. We then fit these spectra as described above, and measured the newly fit profiles. This process was repeated 1000 times in order to obtain a distribution for each of our parameters, and the $68\%$ range is reported as our measurement uncertainty.\\

\section{UV-Based Black Hole Mass Calibration}\label{sec:analysis}

\subsection{Estimating Black Hole Masses}\label{sec:mbh}

\indent In order to perform the analysis discussed in this work, we must first establish H$\beta$-based \mbh estimates (obtained from Paper I), followed by an outline for developing prescriptions for the C~{\sc iv}- and Mg~{\sc ii}-based \mbh estimates.
The initial step is to obtain SE \mbh estimates for each emission line following the virial assumption,

\begin{equation} \label{eq:1}
M_{\rm BH} = \frac{fR_{\rm BELR}\Delta V^{2}}{G},
\end{equation}
where $G$ is the gravitational constant and $f$ is the virial factor which depends on the geometry and orientation of the system and is assumed to be on the order of $\approx1$ \citep[e.g.,][]{HaK14,Yu19}.
The next step is to substitute the continuum luminosity for $R_{\rm BELR}$ according to the $R - L$ relation (see, Section \ref{sec:intro}) as $R_{\rm BELR} \propto L^{0.5}$.

We estimate H$\beta$-based \mbh values by further correcting the $R_{\rm BELR}$ parameter in Equation \ref{eq:1} (hereafter, $R_{\rm H\beta}$) for the source accretion rate, based on the scaling relation presented in \citet{Du19} in the following way
\begin{equation}\label{eq:5}
\log(R_{\rm H\beta}/ {\rm lt - days}) =\alpha +\beta\log\ell_{\rm 44} + \gamma\mathcal{R}_{\rm Fe}
\end{equation}
where $\ell_{\rm 44} = L_{\rm 5100}/10^{44}$ erg s$^{-1}$, $\alpha = 1.65 \pm 0.06$, $\beta = 0.45 \pm 0.03$, $\gamma = -0.35 \pm 0.08$, and $\mathcal{R}_{\rm Fe}$ is an indicator of the strength of the Fe~{\sc ii} emission defined as the ratio of the flux ($F$) or EW between Fe~{\sc ii} \citep[in the $4434$-$4684$\AA\ rest-frame band;][]{BaG92} and H$\beta$; $\mathcal{R}_{\rm Fe} = F_{\rm Fe II}/F_{\rm H\beta} \approx {\rm EW}_{\rm Fe II}/{\rm EW}_{\rm H\beta}$.
In this work we employ the ratio of EWs to determine $\mathcal{R}_{\rm Fe}$.
For the virial factor in Equation \ref{eq:1}, we adopt $f = 1.5$ and the FWHM as $\Delta V$ for H$\beta$-based \mbh values \citep{Mai22}.
The value of the $f$ factor introduces additional uncertainty, on the order of $\sim$2-3 \citep[e.g.,][]{Mej18}, in our estimation of $M_{\rm BH}$. Our adopted value is consistent with \citet{Yu20} and the emipirical best fit value obtained from the $M - \sigma_{\bigstar}$ correlation \citep[e.g.,][]{Onk04,HaK14,Woo15}.

\begin{figure}
\plotone{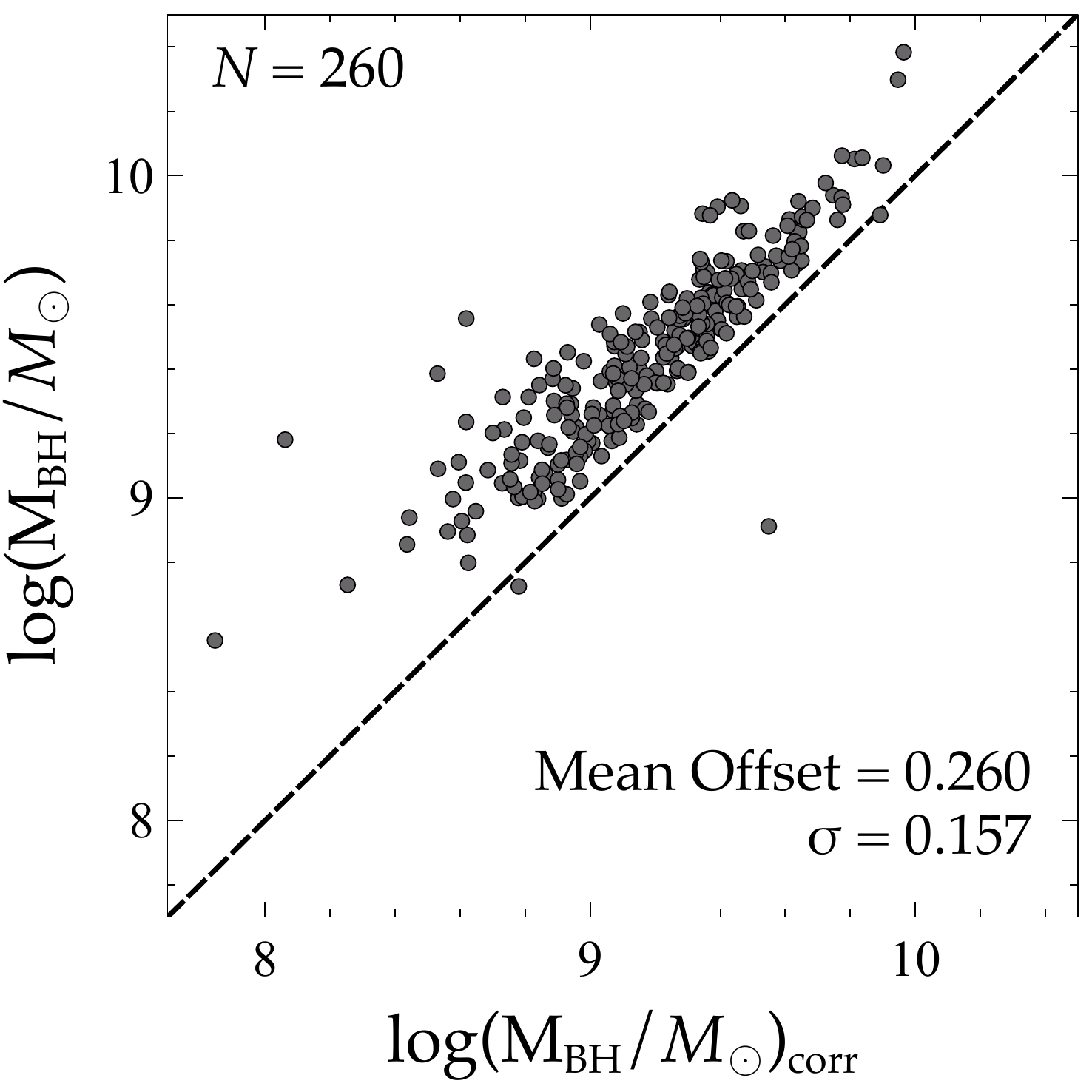}
\caption{The H$\beta$-based \mbh estimates of all 260 quasars from the GNIRS-DQS sample calculated using the VP06 approach (y-axis) and correcting for accretion rate (x-axis). The dashed line represents a one-to-one relationship. This figure shows that H$\beta$-based \mbh esimates that were not corrected for accretion rate are systematically overestimated. \label{fig:hb_comp}}
\end{figure}

\indent \citet{Mai22} have shown that this accretion-rate correction is necessary for adjusting \mbh values that are overestimated by a factor of $\sim 2$ for typical luminous high-redshift quasars.
We compare the accretion rate corrected H$\beta$-based \mbh estimates for our sample to the traditional approach of VP06 which uses the following equation to obtain H$\beta$-based \mbh values:
 
\begin{equation} \label{eq:11}
\begin{split}
\log\bigg( \frac{M_{\rm BH}}{M_{\odot}}\bigg) =0.91 + 2\log\bigg( \frac{{\rm FWHM}_{\rm H\beta}}{\rm km~s^{-1}}\bigg)\\ + 0.5\log\bigg( \frac{\lambda L_{\lambda}(5100  \angstrom)}{10^{44} {\rm erg~s}^{-1}} \bigg),
\end{split}
\end{equation}
utilizing a virial factor on the order of unity.
Figure \ref{fig:hb_comp} presents the H$\beta$-based \mbh masses for our sample, based on the relation of VP06 against our accretion-rate-corrected values.
We find that the masses, computed according to the VP06 approach, are systematically overestimated by 0.26 dex.
This result is consistent with the findings in \citet{Mai22}.

\indent Given that correcting for accretion-rate is necessary for accurate \mbh estimates, we explore whether additional accretion rate based corrections would further improve \mbh estimates for rest-frame UV emission lines.
To accomplish this, we introduce a term into our UV-based \mbh estimates that includes the C~{\sc iv} EW, as this parameter has been shown to be generally anti-correlated with the quasar's accretion rate \citep[e.g.,][]{BaL04,SaL15}. 
Another C~{\sc iv} observable property that is known to be related to the accretion rate is the emission line blueshift with respect to a source systemic redshift \citep[e.g.,][hereafter, Paper III]{BaL05,Ha23}. However, this property cannot be measured reliably when a corresponding indicator of $z_{\rm sys}$ (e.g., the [O~{\sc iii}]~$\lambda$ 5007 emission line) is unavailable. Since our prescriptions for obtaining UV-based \mbh estimates are not restricted to the availability of such indicators, we do not introduce an additional accretion-rate correction term based on C~{\sc iv} blueshift.
%
%

Following Equation \ref{eq:1}, assuming $R_{\rm BELR} \propto L^{0.5}$, with the addition of a C~{\sc iv} EW term, we derive our C~{\sc iv}-based \mbh estimates as

\begin{equation} \label{eq:6}
\begin{split}
\log\bigg( \frac{M_{\rm BH}}{M_{\odot}}\bigg) = 2\log\bigg( \frac{\Delta V}{10^{3} \rm ~km~s^{-1}}\bigg) + 0.5\log\bigg( \frac{\lambda L_{\lambda}(1350  \angstrom)}{10^{44} {\rm erg~s}^{-1}} \bigg) \\ +~a + b\log\bigg(\frac{\rm EW_{\rm C IV}}{\angstrom}\bigg).
\end{split}
\end{equation}
The coefficients $a$ and $b$ were determined from a linear-regression analysis to the calibration set of ($R_{\rm Fe II}$ corrected) H$\beta$-based \mbh estimates.
By design, we allow $a$ and $b$ to freely vary during the regression analysis, resulting in a zero mean offset between the C~{\sc iv}-based and H$\beta$-based \mbh estimates.

The linear-regression was performed such that the difference between our C~{\sc iv}-based \mbh values and the H$\beta$-based \mbh values was minimized.
Specifically, we subtracted the first two terms in Equation \ref{eq:6} from the derived H$\beta$-based \mbh estimates and fit the remaining coefficients, $a$ and $b$, to this difference.
This was accomplished utilizing the {\sc regstats} function in the \hbox{Statistics Toolbox 11.4} of \hbox{MATLAB 9.5.}
As the errors associated with SE \mbh values are large (on the order of 0.5-0.6 dex and 0.7 dex for relative and absolute uncertainty, respectively; see, Section \ref{sec:discussion}), we did not include the errors as part of the linear-regression.
Despite this, we also employed the {\sc linmix\_err} algorithm \citep{Kel07} where we adopted a 0.5 dex uncertainty to have a basis of comparison for our regression, and found the results were generally consistent.
The uncertainty of the coefficients, presented in our equations below, stem directly from the linear fit.

\begin{figure}
\plotone{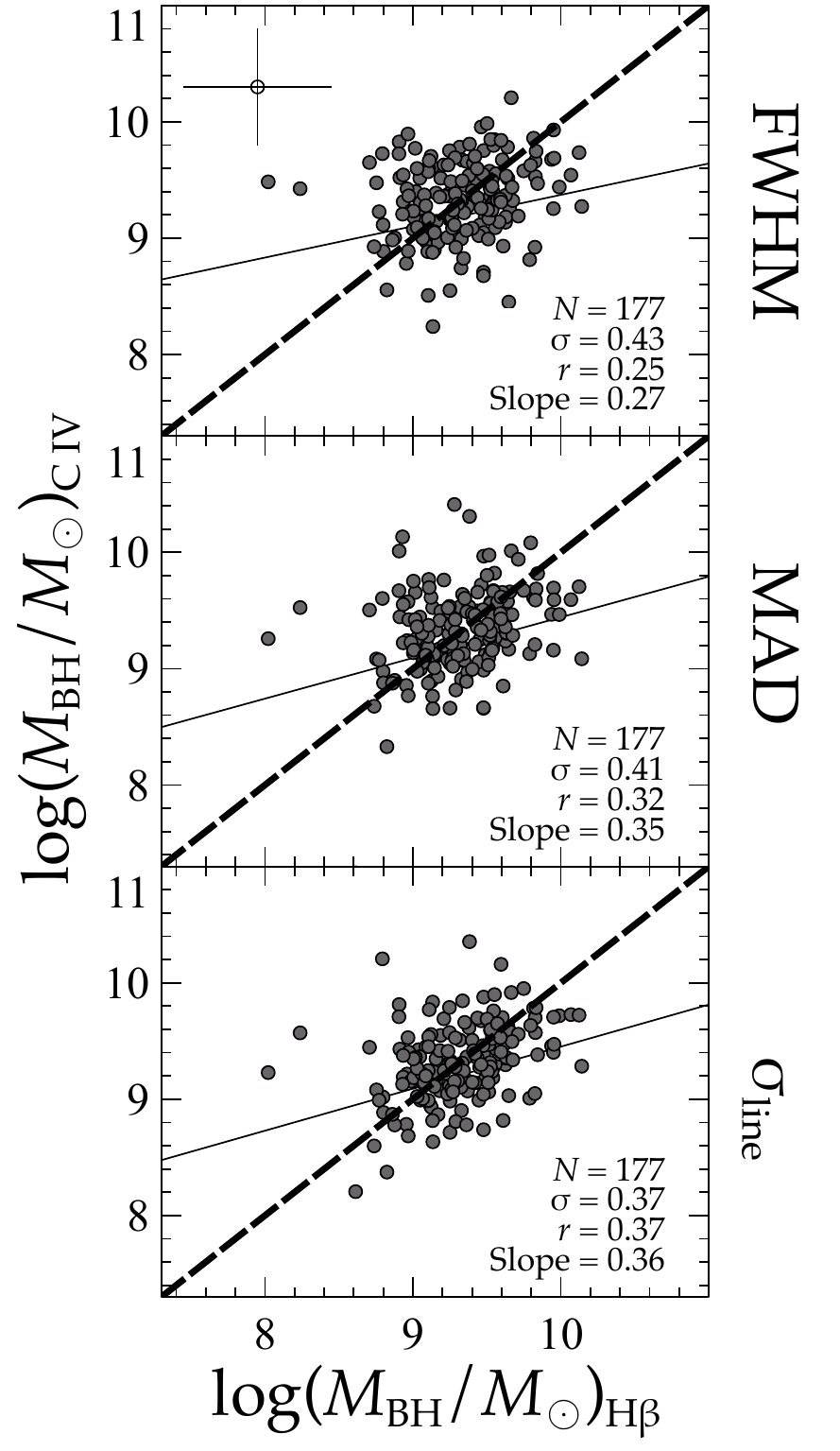}
\caption{The calibrated C~{\sc iv}-based \mbh estimates using the three velocity width parameters, discussed in Section \ref{sec:mbh}, against the calibration set of H$\beta$-based \mbh estimates. The dashed line in each panel represents a one-to-one relationship and the thin solid line in each panel represents the best linear fit to the data. The Pearson correlation coefficient ($r$) and the slope of the best-fit line are provided in each panel. Notably, using $\sigma_{\rm line}$ as the velocity width parameter provides the most precise C~{\sc iv}-based \mbh estimates with respect to the H$\beta$-based \mbh estimates. Additionally, using $\sigma_{\rm line}$ as the velocity width parameter leads to the largest Pearson correlation coefficient and steepest slope of the best fit relation. Typical uncertainty of 0.5 dex on the \mbh values is displayed in the top panel for reference. \label{fig:c4_tot}}
\end{figure}

\indent Our next step is to focus on \mbh estimates that utilize the Mg~{\sc ii} emission line.
Unlike the case for C~{\sc iv} above, we calibrate our Mg~{\sc ii}-based \mbh estimates in two separate runs using the following equation,

\begin{equation} \label{eq:7}
\begin{split}
\log\bigg( \frac{M_{\rm BH}}{M_{\odot}}\bigg) = 2\log\bigg( \frac{\Delta V}{10^{3} \rm ~km~s^{-1}}\bigg) + 0.5\log\bigg( \frac{\lambda L_{\lambda}(3000  \angstrom)}{10^{44} {\rm erg~s}^{-1}} \bigg) \\ +~c + d\log\bigg(\frac{\rm EW_{\rm C IV}}{\angstrom}\bigg),
\end{split}
\end{equation}
where $\Delta V$ is the velocity width of Mg~{\sc ii}; the Mg~{\sc ii} lines were measured from a combination of the SDSS and GNIRS spectra of the sources as described below.
The coefficients $c$ and $d$ were determined differently in each run through a linear-regression analysis to the calibration set of H$\beta$-based \mbh estimates.
The first run set the coefficient $d$ to $0$ in order to provide a prescription that only used the Mg~{\sc ii} emission line while allowing $c$ to be a free parameter.
For this run we did not need any C~{\sc iv} measurements, allowing us to use all of the Mg~{\sc ii} measurements in each subsample (see, Section \ref{sec:sample}).
The second run allowed both $c$ and $d$ to vary freely during the regression.
This run required C~{\sc iv} measurements, reducing our Mg~{\sc ii} sample as described in Section \ref{sec:sample}.
In both runs, we used the same type of linear-regression as discussed for the C~{\sc iv} analysis.

\indent Given the considerably lower S/N ratio of the GNIRS spectra at $\lambda \lesssim 1.2$ $\mu$m (M21), we split the analysis utilizing the Mg~{\sc ii} line measured from the GNIRS spectra into three different parts based on source redshift (see Section \ref{sec:sample}).
In addition to these subsamples, we analyzed the total of 160 and 225 sources for the subsample including all Mg~{\sc ii} measurements (whether from SDSS or GNIRS) with and without C~{\sc iv}, respectively.
For the subsample of 53 sources that have Mg~{\sc ii} measurements available in both the GNIRS and SDSS spectra, the average of these measurements was used in the regression analyses (see Section \ref{sec:mg2_mg2}).
\begin{figure*}
\plotone{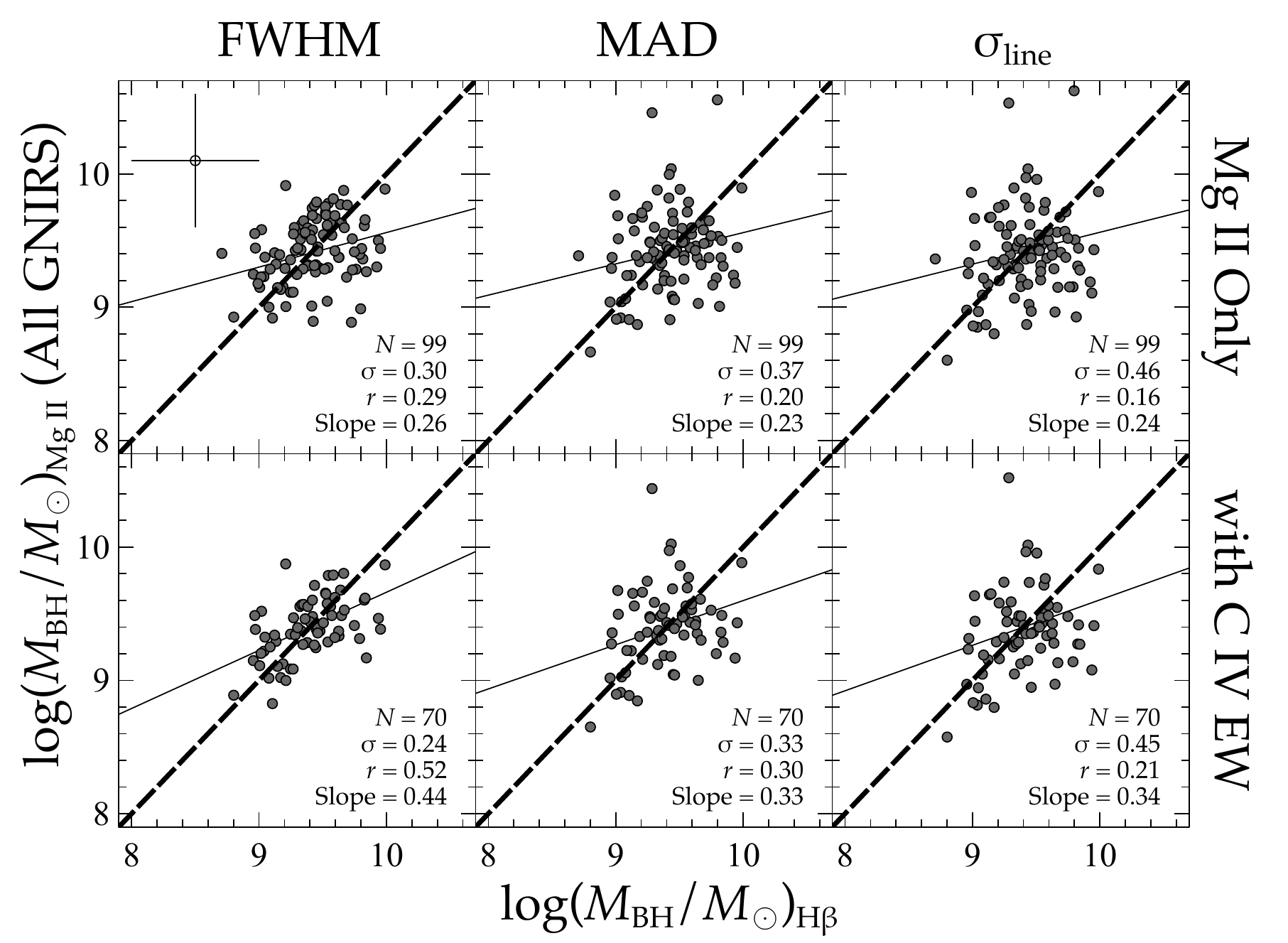}
\caption{Calibrated Mg~{\sc ii}-based \mbh estimates using the three velocity width parameters against the H$\beta$-based \mbh estimates; the bottom panels present the results when adding EW(C~{\sc iv}) to the analysis as discussed in Section \ref{sec:mbh}. The symbols are the same as in Figure \ref{fig:c4_tot}. For all the Mg~{\sc ii}-based \mbh estimates, using the FWHM as the velocity width parameter provided the most precise results when compared to the H$\beta$-based \mbh estimates. For all velocity width parameters, the inclusion of the EW(C~{\sc iv}) parameter, improves the precision of the relation, demonstrated by a reduction in the scatter and an increase in the correlation coefficient in each case. Typical uncertainty of 0.5 dex on the \mbh values is displayed in the top left panel for reference.\label{fig:mg2}}
\end{figure*}
\begin{figure*}
\plotone{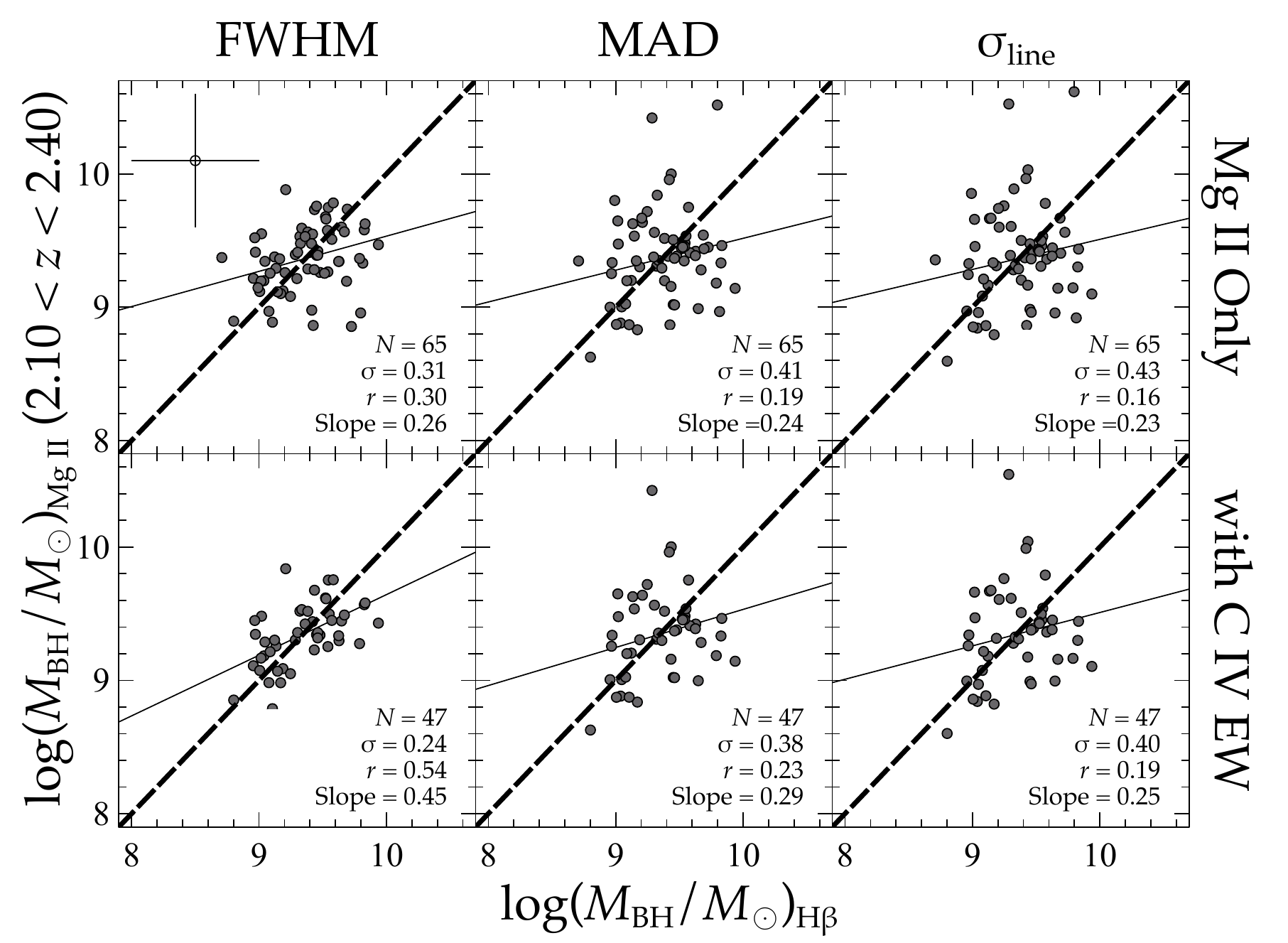}
\caption{Same as Figure \ref{fig:mg2} but for the subset of sources in the range \hbox{$2.10 \lesssim z \lesssim 2.40$}. As observed for the entire redshift range (Figure \ref{fig:mg2}), the FWHM of Mg~{\sc ii} is the most reliable velocity width parameter and the inclusion of the C~{\sc iv} EW helped improve the precision of the Mg~{\sc ii}-based \mbh estimates with respect to those obtained from H$\beta$. \label{fig:mg2_z2}}
\end{figure*}
\begin{figure*}
\plotone{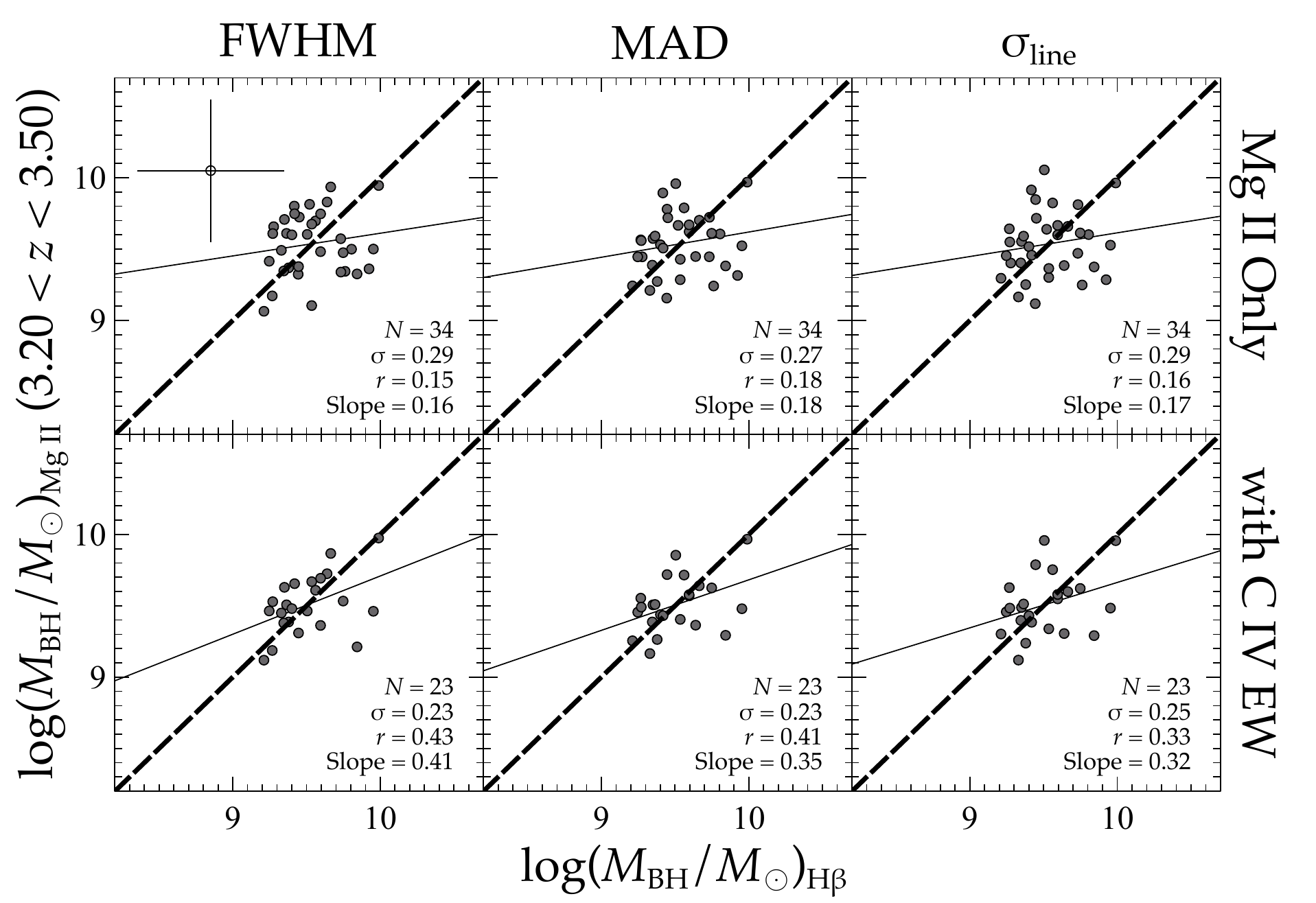}
\caption{Same as Figure \ref{fig:mg2} but for the subset of sources in the range \hbox{$3.20 \lesssim z \lesssim 3.50$}. In this subset of sources the most reliable velocity width parameter for deriving Mg~{\sc ii} only-based \mbh estimates is the MAD instead of the FWHM. This is determined from evaluating the standard deviations and $r$ in each panel. This disparity suggests the importance of expanding the sample of quasars that lie in this redshift range. As we find for the entire redshift range, the inclusion of the EW of C~{\sc iv} (bottom panels) improves the precision of these Mg~{\sc ii}-based \mbh estimates.\label{fig:mg2_z3}}
\end{figure*}
\begin{figure*}
\plotone{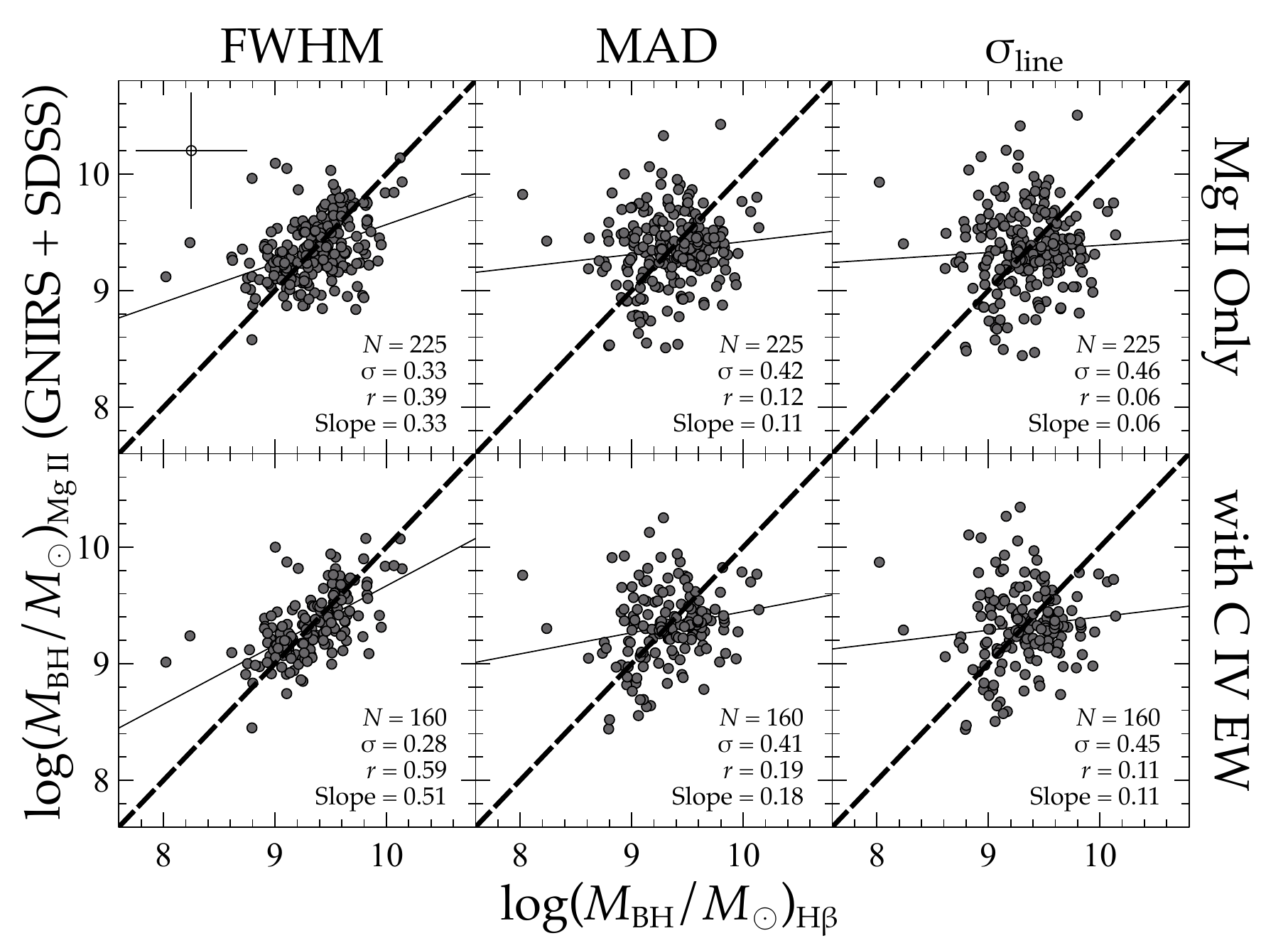}
\caption{Same as Figure \ref{fig:mg2} but for the source sample having Mg~{\sc ii} measurements taken from GNIRS-DQS and/or SDSS. From evaluating the standard deviations and Pearson correlation coefficients in each panel, we find that using the FWHM as the velocity width parameter in the calculation for Mg~{\sc ii}-based \mbh estimates provides the most precise \mbh estimates with respect to the H$\beta$-based \mbh values. As we find for each Mg~{\sc ii} subsample, the inclusion of the EW of C~{\sc iv} (bottom panels) improves the precision of our Mg~{\sc ii}-based \mbh estimates even more. \label{fig:mg2_both}}
\end{figure*}

\subsection{Testing Different Velocity Width Parameters}
We substitute the FWHM, MAD, and $\sigma_{\rm line}$ as the velocity width parameter in each of our \mbh estimates in Equations \ref{eq:6} and \ref{eq:7} to further investigate which of these parameters provides \mbh values closest to those obtained from H$\beta$.
In each analysis described above, we calibrate the C~{\sc iv}- and Mg~{\sc ii}-based \mbh estimates to the H$\beta$-based values that use the FWHM for the velocity width of H$\beta$ \citep{Mai22}.
We determined which velocity width parameter was preferred based on the lowest standard deviation, steepest slope of the best-fit relation and largest Pearson correlation coefficient when comparing the resulting UV- and H$\beta$-based \mbh values.
For the C~{\sc iv}-based \mbh estimates, presented in Figure \ref{fig:c4_tot}, $\sigma_{\rm line}$ produced the most precise results when compared to the H$\beta$-based \mbh values.

For each of the Mg~{\sc ii} subsamples described above, we present the calibrated Mg~{\sc ii}-based \mbh estimates in Figures \ref{fig:mg2}, \ref{fig:mg2_z2}, \ref{fig:mg2_z3}, and \ref{fig:mg2_both} both with (bottom panels) and without (top panels) the inclusion of the C~{\sc iv} EW.
Except for the subsample of sources at \hbox{$3.20 \lesssim z \lesssim 3.50$}, all the other Mg~{\sc ii}-based subsamples showed the strongest corrrelation with the H$\beta$-based \mbh estimates when using the FWHM as the velocity width parameter for the Mg~{\sc ii} line.
For the subsample at \hbox{$3.20 \lesssim z \lesssim 3.50$}, we find that using the MAD for the velocity width parameter in \mbh estimates provides the best results when using only the Mg~{\sc ii} emission line (see, Figure \ref{fig:mg2_z3}).
We recognize that this discrepancy may be a result of the limited sample size which may not provide meaningful statistics.
In spite of this, the results from this subsample are considered to be the least uncertain given that Mg~{\sc ii} and H$\beta$ are measured in the same spectrum with the highest S/N ratio possible.
The best fit coefficients stemming from our linear-regression analyses appear in Table \ref{tab:coeff}.
%
%

\subsection{Comparison with Previous Studies}
\indent In order to have a basis of comparison for this work, we provide estimates for the C~{\sc iv}-based \mbh values for our sample using the prescriptions provided in VP06, \citet[][hereafter P17]{Park17}, and \citet[][hereafter C17]{Coa17}. 
VP06, P17, and C17, use the following Equations to determine C~{\sc iv}-based \mbh estimates, respectively,
\begin{equation}
\begin{split}
\log\bigg( \frac{M_{\rm BH}}{M_{\odot}}\bigg) = 6.66 + 2.0\log\bigg(\frac{\rm FWHM_{\rm C IV}}{10^{3} \rm ~km~s^{-1}}\bigg) \\+ 0.53\log\bigg( \frac{\lambda L_{\lambda}(1350  \angstrom)}{10^{44} {\rm erg~s}^{-1}} \bigg),
\end{split}
\end{equation}

\begin{equation}
\begin{split}
\log\bigg( \frac{M_{\rm BH}}{M_{\odot}}\bigg) = 6.73 + 2.0\log\bigg(\frac{\rm \sigma_{\rm line, C IV}}{10^{3} \rm ~km~s^{-1}}\bigg) \\+ 0.43\log\bigg( \frac{\lambda L_{\lambda}(1350  \angstrom)}{10^{44} {\rm erg~s}^{-1}} \bigg),
\end{split}
\end{equation}

\begin{equation}
\begin{split}
\log\bigg( \frac{M_{\rm BH}}{M_{\odot}}\bigg) = 6.71 + 2.0\log\bigg(\frac{\rm FWHM_{\rm C IV, Corr.}}{10^{3}\rm ~km~s^{-1}}\bigg) \\+ 0.53\log\bigg( \frac{\lambda L_{\lambda}(1350  \angstrom)}{10^{44} {\rm erg~s}^{-1}} \bigg).
\end{split}
\end{equation}
VP06 uses the FWHM as the velocity width, while P17 uses $\sigma_{\rm line}$.
C17 uses a velocity width (FWHM$_{\rm C IV, Corr.}$) that has been adjusted by the blueshift of the C~{\sc iv} emission-line peak with respect to the line peak of H$\beta$ \citep[see,][]{Coa17}.
When evaluating these relations alongside the C~{\sc iv}-based prescription derived in this work, we compare them to the H$\beta$-based \mbh estimates using the FWHM as the velocity width parameter, see Section \ref{sec:mbh}.

\begin{figure}
\plotone{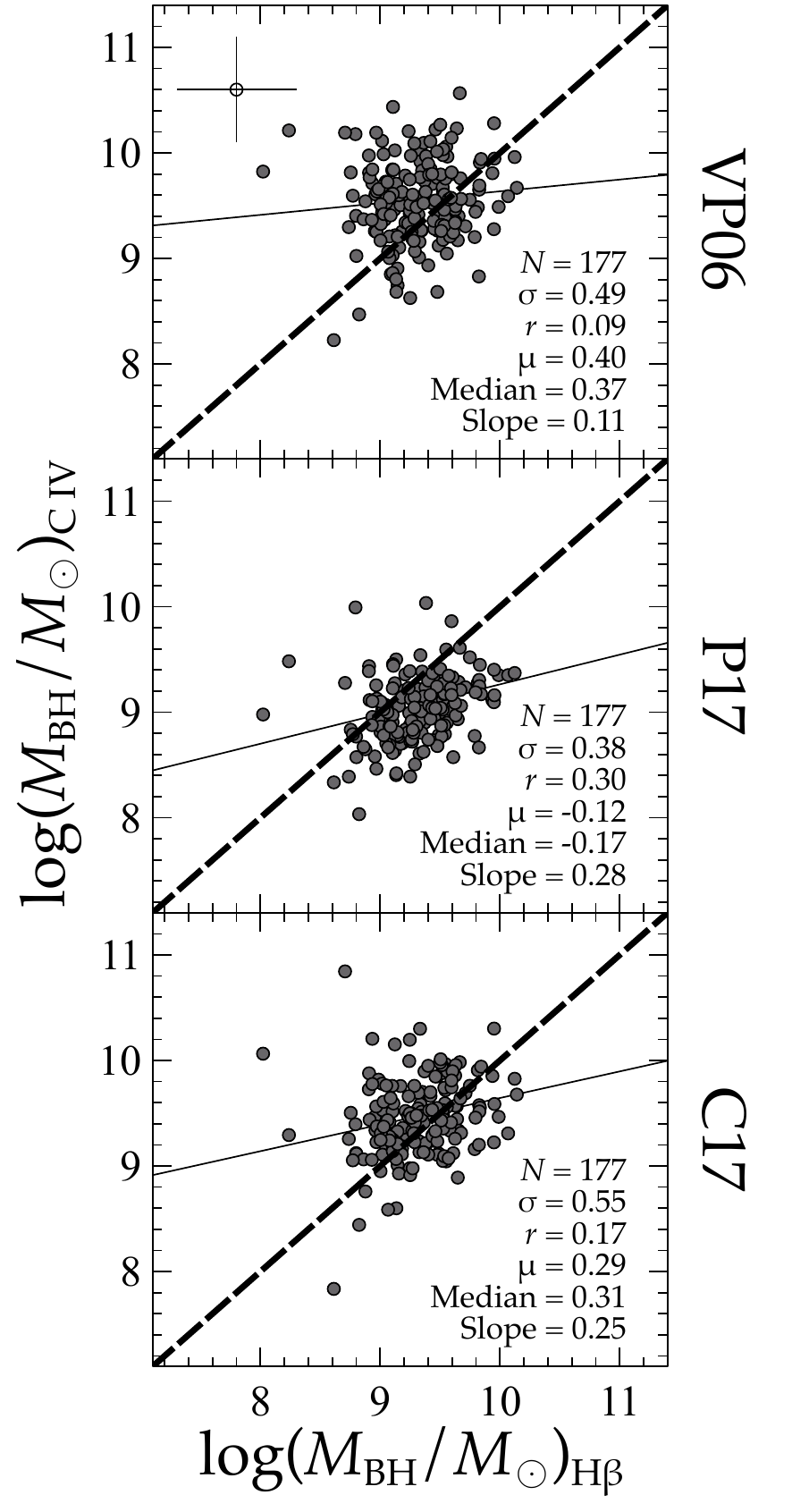}
\caption{C~{\sc iv}-based \mbh estimates of our sample derived through the methodology of, from top to bottom: VP06, P17, and C17 against the H$\beta$-based \mbh estimates. The dashed lines represent one-to-one relationships and the thin solid lines represent the best linear fit to the data in each panel. The most precise C~{\sc iv}-based \mbh values from this work were derived utilizing $\sigma_{\rm line}$ as the velocity width parameter (see the bottom panel of Figure \ref{fig:c4_tot}). Our prescription shows a considerable improvement in the value of the Pearson correlation coefficient, $r$, albeit a modest improvement in the standard deviation, with respect to previous work. Additionally, our prescription corrects the mean offset (the $\mu$ value in each panel) due to considering the accretion rate when estimating H$\beta$-based \mbh values. Typical uncertainty of 0.5 dex on the \mbh values is displayed in the top panel for reference. \label{fig:c4_best}}
\end{figure}

In Figure \ref{fig:c4_best} we present the C~{\sc iv}-based \mbh estimates for our sample based on the prescriptions from the literature.
In comparison, our prescription,

\begin{equation}\label{eq:8}
\begin{split}
\log\bigg( \frac{M_{\rm BH}}{M_{\odot}}\bigg) = (6.299\pm0.169) + 2\log\bigg(\frac{\sigma_{\rm line}}{10^{3} \rm ~km~s^{-1}}\bigg) + \\
0.5\log\bigg( \frac{\lambda L_{\lambda}(1350 \angstrom)}{10^{44} {\rm erg~s}^{-1}} \bigg) + (0.385\pm0.119)\log\bigg(\frac{{\rm EW_{\rm C IV}}}{\angstrom}\bigg),
\end{split}
\end{equation}
which is plotted at the bottom panel of Figure \ref{fig:c4_tot}, provides the smallest scatter, steepest slope of the best-fit relation, largest Pearson correlation coefficient, and, by design, corrects the mean offset\footnote{The mean offset correction accounts for the bias introduced when not considering a source's accretion rate in its H$\beta$-based \mbh value \citep[see,][]{Mai22}.} between previous C~{\sc iv}-based \mbh estimates and H$\beta$-based \mbh value.

\indent To form a basis of comparison for our Mg~{\sc ii}-based \mbh estimates, we followed the prescriptions provided in \citet[][hereafter VO09]{VaO09}, \citet[][hereafter Z15]{Zuo15}, and \citet[][hereafter L20]{Le20}.
VO09, Z15, and L20 use the following Equations to determine Mg~{\sc ii}-based \mbh estimates, respectively,
\begin{equation}\label{eq:m1}
\begin{split}
\log\bigg( \frac{M_{\rm BH}}{M_{\odot}}\bigg) = 0.86 + 2.0\log\bigg(\frac{\rm FWHM_{\rm Mg II}}{\rm  km~s^{-1}}\bigg) \\+ 0.5\log\bigg( \frac{\lambda L_{\lambda}(3000 \angstrom)}{10^{44} {\rm erg~s}^{-1}} \bigg),
\end{split}
\end{equation}

\begin{equation}\label{eq:m2}
\begin{split}
\log\bigg( \frac{M_{\rm BH}}{M_{\odot}}\bigg) = 1.07 + 2.0\log\bigg(\frac{\rm FWHM_{\rm Mg II}}{\rm  km~s^{-1}}\bigg) \\+ 0.48\log\bigg( \frac{\lambda L_{\lambda}(3000 \angstrom)}{10^{44} {\rm erg~s}^{-1}} \bigg),
\end{split}
\end{equation}

\begin{equation}\label{eq:m3}
\begin{split}
\log\bigg( \frac{M_{\rm BH}}{M_{\odot}}\bigg) = 7.00 + 2.0\log\bigg(\frac{\rm FWHM_{\rm Mg II}}{10^{3} \rm  ~km~s^{-1}}\bigg) \\+ 0.5\log\bigg( \frac{\lambda L_{\lambda}(3000 \angstrom)}{10^{44} {\rm erg~s}^{-1}} \bigg).
\end{split}
\end{equation}

\begin{figure}
\plotone{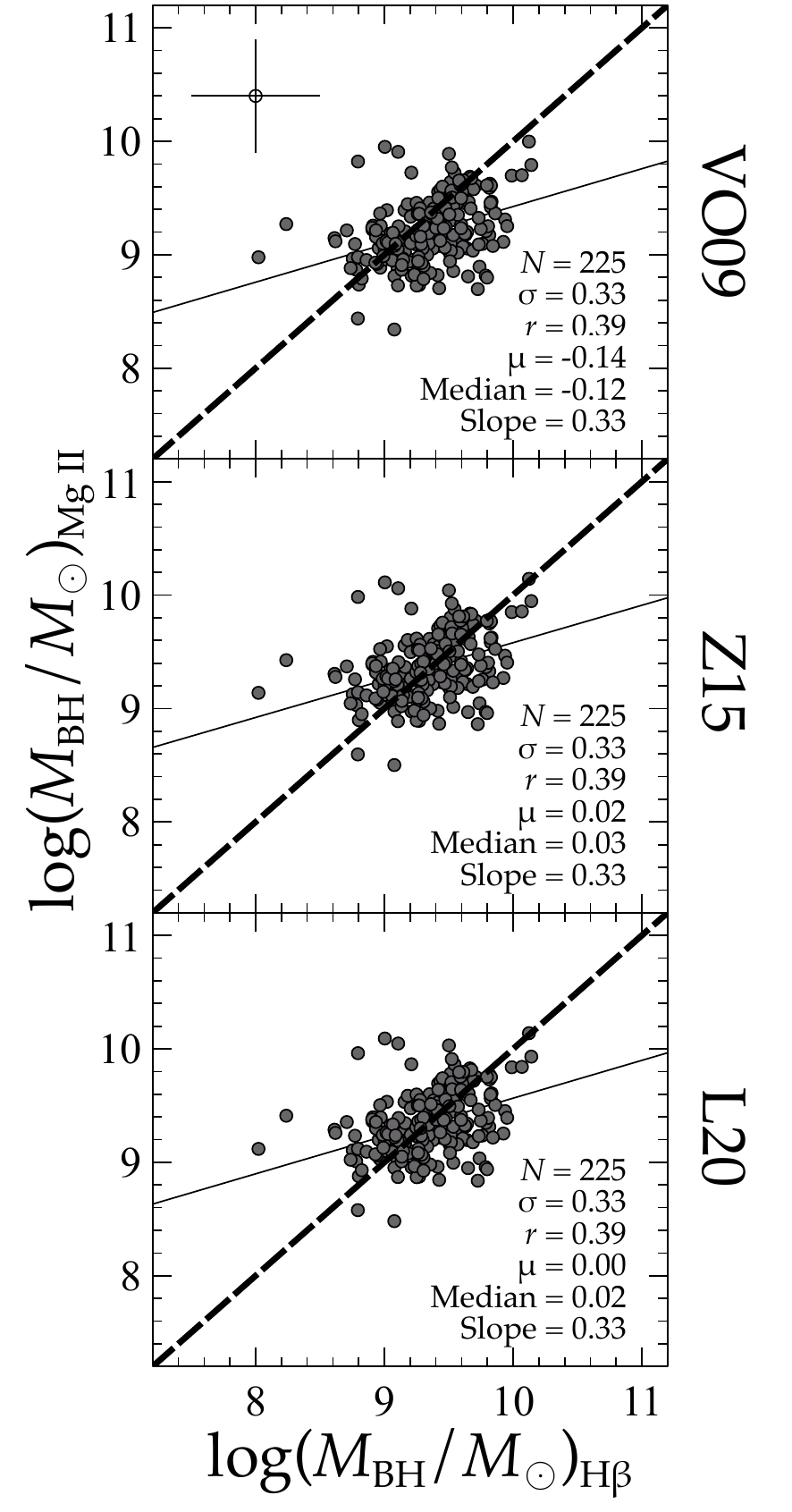}
\caption{Mg~{\sc ii}-based \mbh estimates of our sample derived through the methodology of, from top to bottom, VO09, Z15, and L20 against the H$\beta$-based \mbh estimates. The panels include all Mg~{\sc ii} measurements available in SDSS and/or GNIRS. The dashed line in each panel represents a one-to-one relationship and the thin solid line in each panel represents the best linear fit to the data. We find that our results are consistent with those of previous work when only measuring Mg~{\sc ii}, but are clearly improved with the inclusion of the C~{\sc iv} EW term (see the left most panels of Figure \ref{fig:mg2_both}). Our prescriptions, by design, correct the mean offsets (the $\mu$ value in each panel) between the Mg~{\sc ii}- and H$\beta$-based \mbh values with or without the inclusion of the C~{\sc iv} EW. Typical uncertainty of 0.5 dex on the \mbh values is displayed in the top panel for reference. \label{fig:mg2_best}}
\end{figure}

In Figure \ref{fig:mg2_best}, we present the Mg~{\sc ii}-based \mbh estimates from Equations \ref{eq:m1}, \ref{eq:m2}, and \ref{eq:m3}.
The three panels of Figure \ref{fig:mg2_best} that correspond to these three equations are almost identical to each other given the similarities between these equations.
For comparison, we elect to use the Mg~{\sc ii} subsample that contains SDSS and/or GNIRS measurements as it is the largest and, therefore, provides the most meaningful statistics.
From our comparison, we find that our Mg~{\sc ii}-based \mbh estimates given by,
\begin{equation}\label{eq:9}
\begin{split}
\log\bigg( \frac{M_{\rm BH}}{M_{\odot}}\bigg) = (7.000\pm0.022) + 2\log\bigg(\frac{{\rm FWHM}_{\rm Mg II}}{10^{3} \rm ~km~s^{-1}}\bigg) + \\
0.5\log\bigg( \frac{\lambda L_{\lambda}(3000 \angstrom)}{10^{44} {\rm erg~s}^{-1}} \bigg)
\end{split}
\end{equation}
which is plotted at the top left panel of Figure \ref{fig:mg2_both}, provides results that are consistent with those from the prescriptions of the previous studies except for the mean offset correction stemming from consideration of the accretion rate.
The consistency between Equations \ref{eq:m3} and \ref{eq:9} confirms the results derived in L20.

When the C~{\sc iv} EW is included in the regression analysis for the Mg~{\sc ii}-based \mbh values, we obtain the following prescription (for 160 sources; see, Section \ref{sec:sample}),

\begin{equation}\label{eq:10}
\begin{split}
\log\bigg( \frac{M_{\rm BH}}{M_{\odot}}\bigg) = (6.793\pm0.047) + 2\log\bigg(\frac{{\rm FWHM}_{\rm Mg II}}{10^{3} \rm ~km~s^{-1}}\bigg) \\ + 0.5\log\bigg( \frac{\lambda L_{\lambda} (3000 \angstrom)}{10^{44} {\rm erg~s}^{-1}} \bigg) + \mathbf{(}0.005\pm0.001\mathbf{)}\log\bigg(\frac{{\rm EW}_{\rm C IV}}{\angstrom}\bigg),
\end{split}
\end{equation}
 which is plotted in the bottom left panel of Figure \ref{fig:mg2_both}.
In this case, we see a clear improvement in the scatter, the Pearson correlation coefficient, and slope of the best-fit relation.

We report all the \mbh estimates for the H$\beta$, C~{\sc iv} and Mg~{\sc ii} lines in Table \ref{tab:bhm} where Column (1) provides the SDSS designation of the object, Columns (2), (3), and (4) provide the H$\beta$-based \mbh estimates derived using the FWHM, MAD, and $\sigma_{\rm line}$ as the velocity width, respectively.
Columns (5), (6), and (7) provide C~{\sc iv}-based \mbh estimates derived from VP06, P17, and C17, respectively.
Columns (8), (9), and (10) are the C~{\sc iv}-based estimates derived using the regression analysis for each C~{\sc iv} velocity width parameter, FWHM, MAD, and $\sigma_{\rm line}$, respectively.
We report in columns (11), (12), and (13) the Mg~{\sc ii}-based \mbh estimates derived using the prescriptions of VO09, Z15, and L20.
Lastly, in columns (14), (15), and (16), we report the Mg~{\sc ii}-based \mbh estimates using each of the three Mg~{\sc ii} velocity width parameters, FWHM, MAD, and $\sigma_{\rm line}$, respectively.
For our Mg~{\sc ii}-based \mbh estimates, values are provided with and without the C~{\sc iv} EW term.\\

\subsection{Mg~{\sc ii} Covered by both SDSS and GNIRS Spectra}\label{sec:mg2_mg2}

\begin{figure*}
\plotone{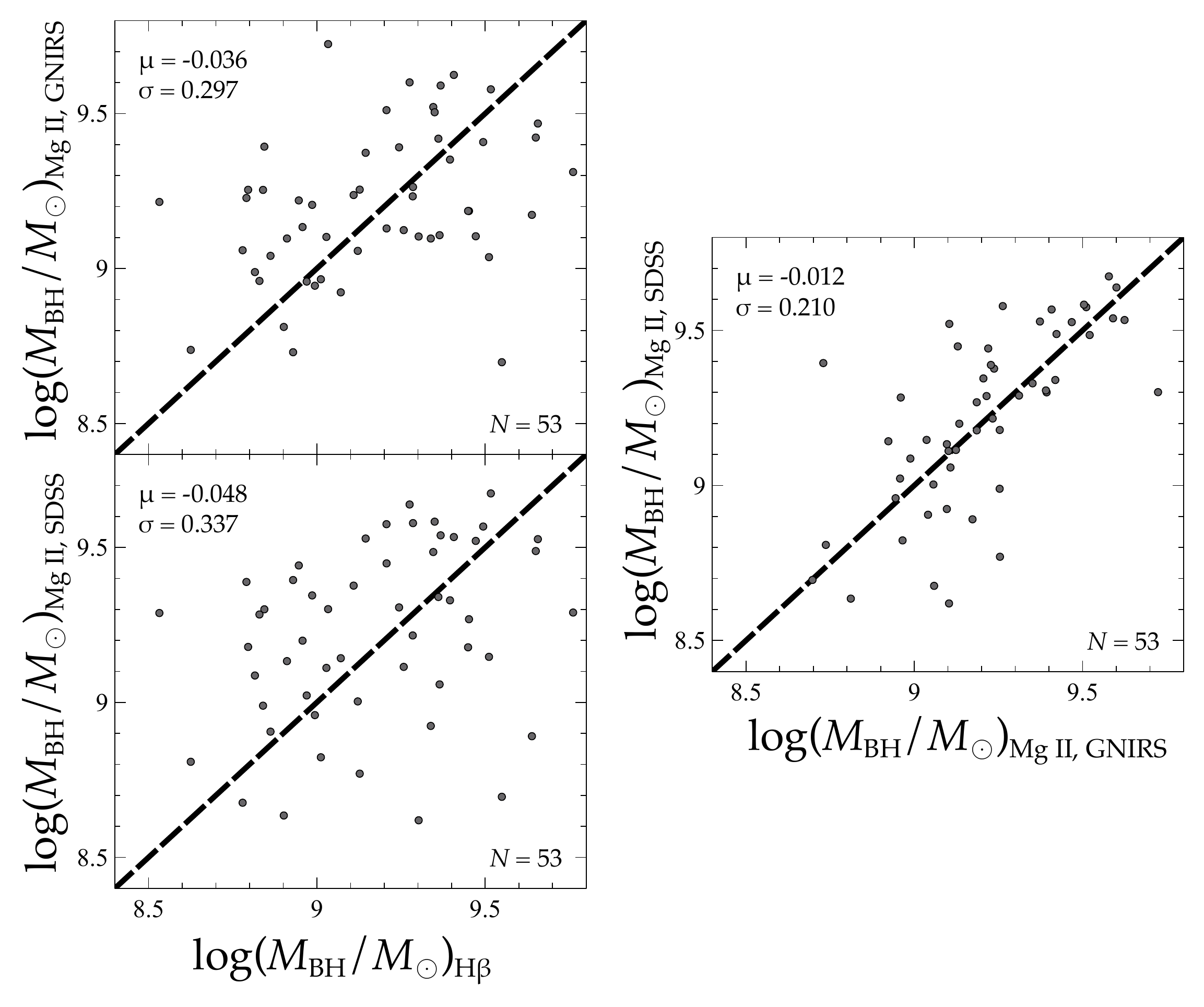}
\caption{The upper leftmost and lower leftmost panel compare the GNIRS-DQS and SDSS, respectively, Mg~{\sc ii}-based \mbh estimates based on the VO09 methodology using the H$\beta$-based masses. The rightmost panel presents the direct comparison of the SDSS- and GNIRS-DQS-based estimates to each other. In each panel, the mean ($\mu$) and standard deviation ($\sigma$) of the residuals with respect to the one-to-one relationship (dashed line) are marked. Overall, we find that the measurements of the Mg~{\sc ii} lines from the GNIRS spectra are consistent with the respective measurements from SDSS. \label{tab:mg2_comp}}
\end{figure*}

\indent For 53 sources from the GNIRS-DQS catalog of Paper I, in the \hbox{$2.10 \lesssim z \lesssim 2.40$} redshift range, we have measurable Mg~{\sc ii} profiles from both GNIRS and SDSS spectra.
In order to confirm consistency across the SDSS and GNIRS spectra, we compare the effects of measuring these spectra in different epochs using different instruments by evaluating the differences in Mg~{\sc ii}-based \mbh estimates stemming from each spectrum.
For consistency, we used the VO09 method for calculating the Mg~{\sc ii}-based \mbh estimates for all measurements in our comparison.
This comparison is presented in Figure \ref{tab:mg2_comp}.
The primary source of the systematic offsets in Figure \ref{tab:mg2_comp} stems from the larger uncertainties of the Mg~{\sc ii} emission line measurements in the GNIRS spectra (see, Paper I).
Overall, we conclude that the two sets of measurements are consistent with each other and the mean offset between the log($M_{\rm BH}$) values is only $-0.012$.

\section{Discussion}\label{sec:discussion}

\indent In this work, we perform calibrations between C~{\sc iv}- and Mg~{\sc ii}-based \mbh estimates and those based on the H$\beta$ line using the largest, homogeneous sample of luminous quasars at high redshift that cover these three emission lines.
The H$\beta$-based \mbh estimates that we calibrate to are accretion-rate-corrected according to the scaling relation presented in \citet{Du19} that involves the optical Fe~{\sc ii} emission.
We show that the inclusion of the C~{\sc iv} EW in our calibrations to these H$\beta$-based \mbh values allow for an additional accretion-rate correction in UV-based \mbh estimates (see also Papers I and III).
The inclusion of this term in our prescriptions leads to UV-based \mbh estimates that are closest to those obtained from H$\beta$.
%

Our results display improvements with respect to similar \mbh calibrations from previous studies that excluded such accretion-rate corrections.
When utilizing $\sigma_{\rm line}$ as the velocity width parameter, we obtain the most robust prescription (Equation \ref{eq:8}) for C~{\sc iv}-based \mbh values, compared with previous studies of this kind.
As shown in the bottom panel of Figure \ref{fig:c4_tot} we reduce the scatter of C~{\sc iv}-based \mbh estimates with respect to those from H$\beta$ by $\sim 24\%$, $\sim 3\%$, and $\sim 33\%$ compared to the prescriptions of VP06, P17, and C17, respectively (see, Figure \ref{fig:c4_best}).
Similarly, the Pearson correlation coefficient between C~{\sc iv}-based and H$\beta$-based \mbh values improves from $0.09$, $0.30$, and $0.17$ to $0.37$, respectively.
The slope of the best-fit relation between C~{\sc iv}-based and H$\beta$-based \mbh values also improves from $0.11$, $0.28$, and $0.25$ to $0.36$, respectively.

We also present a prescription (Equation \ref{eq:9}) for obtaining Mg~{\sc ii}-based \mbh estimates when only the Mg~{\sc ii} line is covered in the spectrum.
This prescription is consistent with the findings of L20, confirming their results.
%
%
It is interesting to note that in the high redshift bin ($3.20 \lesssim z \lesssim 3.50$), the smallest scatter in the Mg~{\sc ii}-based masses (when only the Mg~{\sc ii} line is used) is obtained when the MAD is used as the velocity width.
This scatter, $\sigma=0.27$, is even smaller than the smallest scatter obtained for the entire sample of 225 sources (i.e., when the FWHM is used as the velocity width parameter) by $\sim20\%$ (see, upper panels of Figures \ref{fig:mg2_z3} and \ref{fig:mg2_both}).
%
%
A larger sample of sources in this redshift range is necessary in order to draw firm conclusions as to whether a larger improvement can be achieved.

When we introduce the additional accretion-rate correction factor, in the form of the EW of C~{\sc iv}, we obtain a significantly improved Mg~{\sc ii}-based \mbh value using Equation \ref{eq:10}.
Compared to the Mg~{\sc ii}-based \mbh estimates derived from Equation \ref{eq:9}, this prescription reduces the scatter in the calibration with H$\beta$-based \mbh estimates, by $\sim15\%$.
Similarly, the Pearson correlation coefficient is increased by $\sim51\%$ (see, Figure \ref{fig:mg2_both}).
As we find for the case when only the Mg~{\sc ii} line is available, the scatter in the Mg~{\sc ii}-based masses for the 23 sources in the highest redshift range ($3.20 \lesssim z \lesssim 3.50$) is smaller by $\sim20\%$ than that for the entire sample of 160 sources (see, the bottom left panels of Figures \ref{fig:mg2_z3} and \ref{fig:mg2_both}), emphasizing the need for a larger sample in this redshift range.
%
%
With respect to previous studies discussed throughout this work, our prescriptions, by design, correct the mean offset between UV-based and accretion-rate-corrected H$\beta$-based \mbh estimates.
These corrections are critical, as manifested in Figures \ref{fig:c4_best} and \ref{fig:mg2_best}, where mean offsets of up to 0.40 and 0.14 appear in the $\mu$ values for C~{\sc iv} and Mg~{\sc ii}, respectively.

We note that SE \mbh estimates, in general, have a 0.5-0.6 dex relative uncertainty and 0.7 dex absolute uncertainty (e.g., Table 5, VP06). 
Meanwhile, \mbh measurements that stem from RM campaigns have an inherent uncertainty of 0.3-0.5 dex due to their calibration against the $M - \sigma_{\bigstar}$ relation \citep[e.g.][]{Pet10,Ves11,Shen13,HaK14}, and such observations are quite challenging at high redshift \citep[e.g.,][]{Kas21}.
While not being able to completely bridge the gap between these two approaches, the improvements this work provides to the accuracy and precision of SE UV-based \mbh estimates are considerable.
We find that even when significant outliers are removed from all the \mbh comparisons performed above, the resulting improvements in the scatter of up to $\sim7\%$ do not warrant the removal of otherwise ordinary looking sources from the sample.
Overall, our work shows that when using a large, uniform calibration sample of quasars having coverage of C~{\sc iv}, Mg~{\sc ii}, Fe~{\sc ii} and H$\beta$, and when accounting for accretion rate both in the optical ($\mathcal{R}_{\rm Fe}$) and in the UV (EW(C~{\sc iv})), one can obtain the most reliable prescriptions for obtaining SE UV-based \mbh estimates.
%

\subsection{H$\alpha$-based \mbh values}

The GNIRS-DQS spectral inventory of Paper I also provides measurements for the H$\alpha$ emission line where available.
In order to test the applicability of using this emission line as a \mbh indicator \citep[e.g.,][]{Gre05}, we ran the entire regression analyses presented in this work substituting FWHM(H$\alpha$) for FWHM(H$\beta$).
The standard deviation, mean, and median of the difference between the $\log(M_{\rm BH})$ estimates stemming from these two emission lines were $0.149$, $0.114$, and $0.105$, respectively.
We therefore conclude that the results based on H$\alpha$ are roughly consistent with those obtained from H$\beta$, thereby confirming the applicability of using H$\alpha$ to estimate \mbh values in quasars.

%

\section{Conclusions}\label{sec:conclusions}

\indent We provide prescriptions for reliable rest-frame UV-based \mbh estimates with respect to \mbh estimates obtained from the H$\beta$ line.
Utilizing the GNIRS-DQS catalog (Paper I), we calibrate SE C~{\sc iv}- and Mg~{\sc ii}-based \mbh estimates to H$\beta$-based \mbh estimates using a linear regression analysis that includes two basic accretion-rate observable indicators: the relative strength of the optical Fe~{\sc ii} emission with respect to H$\beta$ and the EW of the C~{\sc iv} emission line. 
We also investigate the use of different velocity width parameters for the C~{\sc iv}- and Mg~{\sc ii}-based \mbh estimates and compare our results with previous studies. We summarize our main results as follows:

\begin{enumerate}
\item The H$\beta$-based \mbh estimates in this work are overestimated by a factor of $\sim2$ when the relative strength of the optical Fe~{\sc ii} emission is not taken into account, consistent with the results of \citet{Mai22}. All of the \mbh prescriptions throughout this work take that correction into account. 

\item The inclusion of the C~{\sc iv} EW in our prescriptions considerably improves the precision of UV-based \mbh estimates. With respect to previous studies, our most reliable UV-based \mbh values reduce the scatter by $\sim15\%$ when compared to H$\beta$-based values.

\item The preferred velocity width parameters for estimating \mbh using C~{\sc iv} and Mg~{\sc ii} are $\sigma_{\rm line}$ and FWHM, respectively.

\item Equation \ref{eq:8} presents the prescription for obtaining the most reliable C~{\sc iv}-based \mbh estimates, in the absence of Mg~{\sc ii} coverage. 
Conversely, if the source's spectrum only covers the Mg~{\sc ii} line, the prescription from Equation \ref{eq:9} is preferred.
%
Otherwise, Equation \ref{eq:10} presents the most robust prescription for UV-based \mbh estimates when there is spectral coverage of both C~{\sc iv} and Mg~{\sc ii} emission lines.

\item NIR observations of additional sources at \hbox{$3.20 \lesssim z \lesssim 3.50$} would allow us to test if further significant improvements can be achieved for UV-based \mbh estimates.
Primarily, this redshift range reduces the uncertainty introduced when measuring Mg~{\sc ii} by shifting the emission line redward from the $J$-band.
A larger sample with high quality spectral data at this redshift range may reveal further discrepancies between low and high luminosity objects.

\end{enumerate}

In the coming decade, we expect that millions of high-redshift ($z \gtrsim 0.8$) quasars will have \mbh estimates derived from rest-frame UV emission lines through large spectroscopic surveys, e.g., the Dark Energy Spectroscopic Instrument \citep[DESI,][]{Levi13,DESI} and the 4m Multi-Object Spectroscopic Telescope \citep{deJ12}.
It is therefore crucial to derive the most reliable \mbh estimates for future high-redshift quasar catalogs using the prescriptions provided in this work.

\section*{Acknowledgments}

We gratefully thank the contributions to this work from Yue Shen and Michael A. Strauss. This work is supported by National Science Foundation grants AST-1815281 (C. D., O. S., B. M. M.), AST-1815645 (M. S. B., A. D. M.), and AST-2106990 (W. N. B.). I.A. acknowledges the support from Universidad Nacional de La Plata through grant 11/G153. We thank an anonymous referee for constructive comments that improved this manuscript. This research has made use of the NASA/IPAC Extragalactic Database (NED), which is operated by the Jet Propulsion Laboratory, California Institute of Technology, under contract with the National Aeronautics and Space Administration.

\software{MATLAB and Statistics Toolbox \citep{MATLAB}}

\begin{deluxetable}{lcccccccccccc}
\rotate
\tablenum{1}
\tablecaption{C~{\sc iv} and Mg~{\sc ii} Spectroscopic Measurements \label{tab:c4}}
\tabletypesize{\fontsize{7}{7}\selectfont}
\tablewidth{0pt}
\tablehead{
\colhead{} & \colhead{}  & \colhead{}&
\colhead{C~{\sc iv}} & \colhead{} & \colhead{} & \colhead{}  & \colhead{}&
\colhead{Mg~{\sc ii}} & \colhead{} & \colhead{} & \colhead{} & \colhead{}\\
\colhead{} & \colhead{FWHM}  & \colhead{MAD}&
\colhead{$\sigma_{\rm line}$} & \colhead{EW} & \colhead{$\lambda_{\rm peak}$} & \colhead{FWHM}  & \colhead{MAD}&
\colhead{$\sigma_{\rm line}$} & \colhead{EW} & \colhead{$\lambda_{\rm peak}$} & \colhead{$\log(\lambda L_{1350\angstrom})$} & \colhead{$\log(\lambda L_{3000\angstrom})$}\\
\colhead{Quasar} & \colhead{(\kms)}  & \colhead{(\kms)}&
\colhead{(\kms)} & \colhead{(\AA)} & \colhead{(\AA)} & \colhead{(\kms)}  & \colhead{(\kms)}&
\colhead{(\kms)} & \colhead{(\AA)} & \colhead{(\AA)} & \colhead{(erg s$^{-1}$)} & \colhead{(erg s$^{-1}$)}
}
\decimalcolnumbers
\startdata
SDSS J001018.88+280932.5 & $2517\substack{+  53 \\ -  78}$ & $2274\substack{+  37 \\ -  54}$ & $3158\substack{+  55 \\ -  82}$ & $  61\substack{+   1 \\ -   1}$ & $ 4045\substack{+   0 \\ -   0}$ & \nodata & \nodata & \nodata & \nodata & \nodata & $46.4$ & \nodata \\
SDSS J001249.89+285552.6 & \nodata & \nodata & \nodata & \nodata & \nodata & $4195\substack{+ 188 \\ - 249}$ & $2183\substack{+ 480 \\ - 757}$ & $2956\substack{+ 637 \\ -1017}$ & $  21\substack{+   7 \\ -   9}$ & $11874\substack{+   3 \\ -   5}$ & \nodata & $46.9$\\
SDSS J001355.10-012304.0 & \nodata & \nodata & \nodata & \nodata & \nodata & $2815\substack{+ 344 \\ - 455}$ & $1249\substack{+ 260 \\ - 401}$ & $1595\substack{+ 332 \\ - 515}$ & $  17\substack{+   1 \\ -   2}$ & $12274\substack{+   4 \\ -   5}$ & \nodata & $46.7$\\
SDSS J001453.20+091217.6 & $6487\substack{+ 822 \\ -1227}$ & $3798\substack{+ 910 \\ -1358}$ & $5788\substack{+1383 \\ -2064}$ & $  39\substack{+   3 \\ -   5}$ & $ 5152\substack{+   5 \\ -   7}$ & $ 2999\substack{+   943 \\ -   1248}$ & $ 1833\substack{+   1906 \\ -   1833}$ & $ 2375\substack{+   2580 \\ -   2375}$ & $ 25\substack{+   8 \\ -   11}$ & $ 9374\substack{+   8 \\ -   10}$ & $46.4$ & $46.5$\\
SDSS J001813.30+361058.6 & $6079\substack{+ 197 \\ - 294}$ & $3247\substack{+ 238 \\ - 356}$ & $3861\substack{+ 369 \\ - 550}$ & $  26\substack{+   1 \\ -   2}$ & $ 5116\substack{+   2 \\ -   3}$ & $5129\substack{+ 983 \\ -1301}$ & $3354\substack{+1648 \\ -2632}$ & $4543\substack{+2198 \\ -3511}$ & $  25\substack{+   8 \\ -  11}$ & $ 9303\substack{+  13 \\ -  17}$ & $46.8$ & $46.6$\\
SDSS J001914.46+155555.9 & $4162\substack{+ 215 \\ - 320}$ & $2329\substack{+  81 \\ - 120}$ & $3038\substack{+ 121 \\ - 180}$ & $  45\substack{+   1 \\ -   1}$ & $ 5054\substack{+   1 \\ -   2}$ & $4380\substack{+ 327 \\ - 433}$ & $1628\substack{+ 821 \\ -1235}$ & $2061\substack{+1092 \\ -1643}$ & $  23\substack{+   1 \\ -   1}$ & $ 9141\substack{+   5 \\ -   6}$ & $46.7$ & $46.5$\\
SDSS J002634.46+274015.5 & $5196\substack{+ 739 \\ -1103}$ & $6331\substack{+ 868 \\ -1295}$ & $6701\substack{+1462 \\ -2181}$ & $ 135\substack{+  10 \\ -  15}$ & $ 5023\substack{+   5 \\ -   7}$ & $3158\substack{+ 150 \\ - 198}$ & $1747\substack{+ 645 \\ - 979}$ & $2373\substack{+ 934 \\ -1418}$ & $  36\substack{+   1 \\ -   1}$ & $ 9097\substack{+   2 \\ -   2}$ & $46.2$ & $46.5$ \\
SDSS J003001.11-015743.5 & $6077\substack{+ 265 \\ - 396}$ & $3339\substack{+ 251 \\ - 374}$ & $3719\substack{+ 449 \\ - 669}$ & $  53\substack{+   2 \\ -   3}$ & $ 3995\substack{+   1 \\ -   2}$ & \nodata & \nodata & \nodata & \nodata & \nodata & $45.9$ & \nodata \\
SDSS J003416.61+002241.1 & $4213\substack{+ 107 \\ - 160}$ & $2092\substack{+  43 \\ -  65}$ & $2710\substack{+  66 \\ -  98}$ & $  29\substack{+   0 \\ -   0}$ & $ 4067\substack{+   1 \\ -   1}$ & $ 4141\substack{+   203 \\ -   269}$ & $ 1767\substack{+   233 \\ -   308}$ & $ 2278\substack{+   414 \\ -   548}$ & $ 39\substack{+   2 \\ -   2}$ & $ 7366\substack{+   2 \\ -   2}$ & $46.4$ & $46.4$\\
SDSS J003853.15+333044.3 & $8273\substack{+ 564 \\ - 841}$ & $2485\substack{+ 403 \\ - 602}$ & $3817\substack{+ 593 \\ - 884}$ & $  14\substack{+   1 \\ -   1}$ & $ 5213\substack{+  11 \\ -  17}$ & \nodata & \nodata & \nodata & \nodata & \nodata & $46.3$ & \nodata\\
\enddata
\tablecomments{C~{\sc iv} and Mg~{\sc ii} emission line measurements for the first ten quasars in our sample. The entire table is available online.}
\end{deluxetable}

\begin{deluxetable*}{lccc}
\tablecaption{Regression Coefficients  \label{tab:coeff}}
\tablenum{2} 
\tabletypesize{\fontsize{9.5}{9.5}\selectfont}
\tablehead
{
\colhead{Emission Line} & \colhead{FWHM} & \colhead{MAD} & \colhead{$\sigma_{\rm line}$}
}
\startdata
C~{\sc iv} ($a,b$) & ($5.172\pm0.196, 0.960\pm0.138$) & ($6.727\pm0.187, 0.250\pm0.131$) & \textbf{(6.299$\pm$0.169, 0.385$\pm$0.119)} \\
Mg~{\sc ii} only ($c,d$) & \textbf{(7.000$\pm$0.022, 0)} & ($7.562\pm0.028, 0$) & ($7.309\pm0.031, 0$) \\
Mg~{\sc ii} \& C~{\sc iv} ($c,d$) & \textbf{(6.793$\pm$0.047, 0.005$\pm$0.001)} & ($7.410\pm0.0.068, 0.005\pm0.002$) & ($7.168\pm0.074, 0.004\pm0.002$) 
\enddata
\tablecomments{Resulting regression coefficients from Equations \ref{eq:6} and \ref{eq:7} for each of our velocity width parameters. Bold-faced coefficients are the recommended prescription for each emission line (see, Section \ref{sec:discussion}).}
\end{deluxetable*}

\begin{longrotatetable} 
\begin{deluxetable*}{lccccccccccccccc} 
\tablenum{3} 
\tablecaption{$M_{\rm BH}$ Estimates \label{tab:bhm}} 
\tablewidth{0pt} 
\tablehead{ 
\colhead{} & \colhead{H$\beta$} & \colhead{} & \colhead{} & 
\colhead{C~{\sc iv}} & \colhead{} & \colhead{} &\colhead{} & \colhead{} & \colhead{} &
\colhead{Mg~{\sc ii}} &\colhead{} & \colhead{} &\colhead{} & \colhead{} & \colhead{}\\ 
\colhead{Quasar} & \colhead{FWHM}  & \colhead{MAD}& \colhead{$\sigma_{\rm line}$} & 
\colhead{VP06} & \colhead{P17} & \colhead{C17} & \colhead{FWHM} &\colhead{MAD} &\colhead{$\sigma_{\rm line}$} &
\colhead{VO09} & \colhead{Z15} & \colhead{L20} &\colhead{FWHM\tablenotemark{a}} &\colhead{MAD\tablenotemark{a}} &\colhead{$\sigma_{\rm line}$\tablenotemark{a}} 
} 
\decimalcolnumbers
\startdata
SDSS J001018.88+280932.5 & $ 9.15$ & $8.58$ & $8.82$ & $ 8.74$ & $ 8.77$ & $ 9.01$ & $8.90$ & $9.10$ & $9.20$ & \nodata & \nodata & \nodata & \shortstack{ \nodata \\ \nodata} & \shortstack{ \nodata \\ \nodata} & \shortstack{ \nodata \\ \nodata}\\
SDSS J001249.89+285552.6 & $ 9.42$ & $8.75$ & $8.99$ & \nodata & \nodata & \nodata & \nodata & \nodata & \nodata & $ 9.55$ & $ 9.71$ & $ 9.69$ & \shortstack{ \nodata \\ $ 9.69$} & \shortstack{ \nodata \\ $ 9.69$} & \shortstack{ \nodata \\ $ 9.70$}\\
SDSS J001355.10-012304.0 & $ 9.92$ & $9.22$ & $9.44$ & \nodata & \nodata & \nodata &  \nodata & \nodata & \nodata & $ 9.11$ & $ 9.27$ & $ 9.25$ & \shortstack{ \nodata \\ $ 9.25$} & \shortstack{ \nodata \\ $ 9.11$} & \shortstack{ \nodata \\ $ 9.07$}\\
SDSS J001453.20+091217.6 & $ 9.64$ & $8.70$ & $8.90$ & $ 9.55$ & $ 9.28$ & $ 9.44$ & $9.51$ & $9.47$ & $9.63$ &v$ 9.05$ & $ 9.22$ & $ 9.20$ & \shortstack{$ 9.20$ \\$ 9.20$} & \shortstack{$ 9.36$ \\$ 9.33$} & \shortstack{$ 9.33$ \\$ 9.30$}\\
SDSS J001813.30+361058.6 & $ 9.44$ & $8.61$ & $8.82$ & $ 9.71$ & $ 9.10$ & $ 9.29$ & $9.49$ & $9.50$ & $9.41$ & $ 9.57$ & $ 9.73$ & $ 9.71$ & \shortstack{$ 9.65$ \\$ 9.71$} & \shortstack{$ 9.87$ \\$ 9.91$} & \shortstack{$ 9.89$ \\$ 9.92$}\\
SDSS J001914.46+155555.9 & $ 9.32$ & $8.81$ & $9.08$ & $ 9.30$ & $ 8.83$ & $ 9.38$ & $9.32$ & $9.20$ & $9.22$ & $ 9.37$ & $ 9.53$ & $ 9.51$ & \shortstack{$ 9.54$ \\$ 9.51$} & \shortstack{$ 9.27$ \\$ 9.22$} & \shortstack{$ 9.22$ \\$ 9.17$}\\
SDSS J002634.46+274015.5 & $ 9.48$ & $8.86$ & $9.09$ & $ 9.26$ & $ 9.33$ & $ 9.48$ & $9.75$ & $9.97$ & $9.88$ & $9.10$ & $ 9.26$ & $ 9.24$ & \shortstack{$ 9.76$ \\$ 9.24$} & \shortstack{$ 9.75$ \\$ 9.29$} & \shortstack{$ 9.74$ \\$ 9.30$}\\
SDSS J003001.11-015743.5 & $ 9.18$ & $8.50$ & $8.71$ & $ 9.25$ & $ 8.70$ & $ 9.17$ & $9.36$ & $9.17$ & $9.07$ & \nodata & \nodata & \nodata & \shortstack{ \nodata \\ \nodata} & \shortstack{ \nodata \\ \nodata} & \shortstack{ \nodata \\ \nodata}\\
SDSS J003416.61+002241.1 & $ 9.33$ & $8.71$ & $8.96$ & $ 9.16$ & $ 8.61$ & $ 9.13$ & $9.00$ & $8.91$ & $8.90$ & $ 9.27$ & $ 9.44$ & $ 9.41$ & \shortstack{$ 9.36$ \\$ 9.41$} & \shortstack{$ 9.21$ \\$ 9.23$} & \shortstack{$ 9.18$ \\$ 9.20$}\\
SDSS J003853.15+333044.3 & $ 9.37$ & $8.60$ & $8.83$ & $ 9.73$ & $ 8.90$ & $ 9.90$ & $9.27$ & $8.97$ & $9.07$ & \nodata & \nodata & \nodata &  \shortstack{ \nodata \\ \nodata} &  \shortstack{ \nodata \\ \nodata} &  \shortstack{ \nodata \\ \nodata}\\
\enddata
\tablenotetext{a}{log($M_{\rm BH}$/$M_{\odot}$) estimates derived with (top row) and without (bottom row) the inclusion of the C~{\sc iv} EW, where available.}
\tablecomments{Data for 10 sources are shown. The entire table is available online.}
\end{deluxetable*}
\end{longrotatetable}

\end{document}